\documentclass[journal,twoside,web]{ieeecolor2}
\usepackage{generic}
\usepackage{cite}
\usepackage{amsmath,amssymb,amsfonts}
\usepackage{algorithmic}
\usepackage{graphicx}
\usepackage{algorithm,algorithmic}
\usepackage{hyperref}
\usepackage{colortbl}
\usepackage{textcomp}
\usepackage{array}

\newcolumntype{?}{!{\vrule width 2pt}}

\newcommand{\bigasterisk}{\textsuperscript{\textbf{\Huge *}}}
\def\BibTeX{{\rm B\kern-.05em{\sc i\kern-.025em b}\kern-.08em
    T\kern-.1667em\lower.7ex\hbox{E}\kern-.125emX}}

\begin{document}
\title{DINOMotion: advanced robust tissue motion tracking with DINOv2 in 2D-Cine MRI-guided radiotherapy}
\author{Soorena Salari, Catherine Spino, Laurie-Anne Pharand, Fabienne Lathuiliere, Hassan Rivaz \IEEEmembership{Senior Member, IEEE}, Silvain Beriault, Yiming Xiao \IEEEmembership{Senior Member, IEEE}
\thanks{Soorena Salari and Yiming Xiao are with the Department of Computer Science and Software Engineering, Concordia University, Montreal, Canada (e-mail: soorena.salari@concordia.ca and yiming.xiao@concordia.ca).}
\thanks{Hassan Rivaz is with the Department of Electrical and Computer Engineering, Concordia
University, Montreal, Canada (e-mail: hassan.rivaz@concordia.ca).}
\thanks{Catherine Spino, Laurie-Anne Pharand, and Fabienne Lathuiliere were with the Elekta Ltd., Montreal, Canada (e-mail: catherine.spino@polymtl.ca, laurie-anne.pharand@polymtl.ca, and fabienne.lathuiliere@gmail.com).}
\thanks{Silvain Beriault was with Elekta Ltd., Montreal, Canada. He is now with Diagnos Medical Systems, Montreal, QC, Canada (e-mail: silvain.beriault@gmail.com).}
}

\maketitle

\begin{abstract}
Accurate tissue motion tracking is critical to ensure treatment outcome and safety in 2D-Cine MRI-guided radiotherapy. This is typically achieved by registration of sequential images, but existing methods often face challenges with large misalignments and lack of interpretability. In this paper, we introduce DINOMotion, a novel deep learning framework based on DINOv2 with Low-Rank Adaptation (LoRA) layers for robust, efficient, and interpretable motion tracking. DINOMotion automatically detects corresponding landmarks to derive optimal image registration, enhancing interpretability by providing explicit visual correspondences between sequential images. The integration of LoRA layers reduces trainable parameters, improving training efficiency, while DINOv2’s powerful feature representations offer robustness against large misalignments. Unlike iterative optimization-based methods, DINOMotion directly computes image registration at test time. Our experiments on volunteer and patient datasets demonstrate its effectiveness in estimating both linear and nonlinear transformations, achieving Dice scores of 92.07\% for the kidney, 90.90\% for the liver, and 95.23\% for the lung, with corresponding Hausdorff distances of 5.47 mm, 8.31 mm, and 6.72 mm, respectively. DINOMotion processes each scan in approximately 30ms and consistently outperforms state-of-the-art methods, particularly in handling large misalignments. These results highlight its potential as a robust and interpretable solution for real-time motion tracking in 2D-Cine MRI-guided radiotherapy.
\end{abstract}

\begin{IEEEkeywords}
Cine MRI, Deep Learning, MRI-Guided Radiotherapy, Motion Tracking, Registration, Foundation Model 
\end{IEEEkeywords}

\section{Introduction}
\label{sec:introduction}
New advancements in medical imaging technology have enabled the commercialization of MRI-guided radiotherapy, potentially setting a new standard for image-guided treatments in clinical practice \cite{raaymakers2017first, raaymakers2009integrating}. This innovation combines a linear accelerator for external beam radiotherapy with a magnetic resonance imaging (MRI) scanner, allowing real-time visualization of tumors and surrounding organs during treatment \cite{hunt2018adaptive,corradini2019mr}. The real-time acquisition of 2D-cine MRI provides continuous imaging, capturing various types of motions, such as slow drifts and sudden shifts in both the target and surrounding organs \cite{cusumano2018predicting}. This capability enables the monitoring of moving tumors and organs-at-risk, greatly enhancing the accuracy, effectiveness, and safety of treatment. In clinical settings, accurate motion tracking of these areas is essential for precise radiation delivery. This is often achieved by comparing images acquired at different time points through image registration. By automatically aligning sequential images to a reference template, patient movements are accurately detected and quantified, ensuring that radiation precisely targets the tumor without harming healthy tissues. This real-time motion tracking is crucial for adapting radiation delivery \cite{jassar2023real}, leading to improved clinical outcomes. While 2D-cine MRI provides critical information for real-time motion, effective motion management requires more than just imaging; it also necessitates fast, reliable, and robust localization techniques. In this work, we focus on enhancing motion tracking through advanced localization-based image registration methods, enabling continuous and automated tracking of patient movements for immediate adjustments to radiation delivery.

To allow efficient and precise tissue motion tracking in MRI-guided radiotherapy, automatic 2D-Cine MRI registration algorithms have been frequently used. Mazur et al. \cite{mazur2016sift} were among the first to use cine MRI from the 0.35T MRI-Linac for automated tracking. They employed the scale-invariant feature transform (SIFT) with deformable spatial pyramid matching for motion tracking. Despite good tracking accuracy, the method was too slow (250 $ms$ per frame) for clinical use. With the rise of deep learning (DL) algorithms, recent studies have started to adopt DL-based techniques for improved motion tracking, with most prior methods relying on convolutional neural networks (CNNs). For example, Terpstra et al. \cite{terpstra2020deep} showed that a CNN could surpass the performance of a traditional optical flow algorithm in estimating 2D motions for abdominal cancer patients. They later expanded their approach to a 3D lung dataset by employing a hierarchical CNN at multiple resolutions. Also, Frueh et al. \cite{frueh2022self} demonstrated that a ResNet-18 model with self-supervised learning can surpass optical flow approaches for cardiac and abdominal cine MRI. Hunt et al. \cite{hunt2023fast} adapted VoxelMorph \cite{balakrishnan2019voxelmorph} for quickly generating deformation vector fields (DVFs), demonstrating enhanced performance compared to the traditional counterparts. In MRI-guided radiotherapy, sudden large motions occur due to a variety of anomalies (e.g., sudden swallowing, sudden target shift, etc). However, DL models for motion tracking, especially CNN-based ones, often fail when encountering image pairs with large misalignments because they have difficulty capturing long-range dependencies and correspondences due to their limited receptive field. Many deep registration techniques concatenate moving and fixed images across different channels to create the input tensor. This approach faces challenges with large displacements, as the initial fine-scale convolution layers of the network attempt to extract features from unrelated regions of the image pair \cite{hachicha2023robust,wang2023robust,evan2022keymorph}. Finally, these models lack interpretability as they function as black-box systems that provide transformation parameters or deformation fields without providing insights into the factors influencing the alignment. \textcolor{black}{Although post-hoc interpretability methods, such as saliency maps \cite{simonyan2013deep} and Grad-CAM \cite{selvaraju2017grad, selvaraju2020grad} exist for CNNs, they are typically indirect, gradient-based, and often difficult to validate, particularly for image registration tasks that require precise, pairwise anatomical correspondence rather than class-specific activations.} Attempted to mitigate this, Wang et al. \cite{wang2023robust,evan2022keymorph} introduced KeyMorph, a DL-based framework that detects keypoint pairs for image registration. However, it relies on a CNN backbone and is specifically designed for 3D brain MRI registration. Tailored DL solutions that enable interpretable registration for 2D Cine MRI-guided radiotherapy would be highly beneficial.

For deep learning with medical images, obtaining large amounts of data for tasks like image registration is often challenging due to patient privacy concerns and the high cost of expert annotations. To address this challenge, we utilize the recent foundational model DINOv2 \cite{oquab2023DINOv2} in our framework. DINOv2, a self-supervised model designed for various vision tasks, learns comprehensive visual features without requiring labeled data. This capability makes DINOv2 particularly suitable for medical applications where data scarcity is a significant issue. By integrating specialized decoders, DINOv2 can be easily adapted to various downstream applications, such as motion tracking/image registration. We aim to fine-tune the DINOv2 encoder, leveraging its extensively pre-trained parameters to improve performance in our designated task. Inspired by previous research \cite{wang2023robust,evan2022keymorph}, the key idea behind our approach is that learned landmarks can be used to compute the optimal transformation directly. These homologous landmarks are extracted using a DINOv2-based model from pairs of moving and template scans. Most importantly, they are optimized specifically for registration and motion tracking, without requiring ground-truth annotations. By selecting a transformation with differentiable parameters based on the learned landmarks, we enable end-to-end training of the motion-tracking pipeline, allowing the model to optimize both landmark discovery and transformation simultaneously. In addition, a visual comparison of the landmark pair alignment offers an intuitive interpretation of the registration outcomes. In particular, our main contributions and findings are: 

\begin{itemize}

  \item We are the first to adopt the DINOv2 foundation model to investigate its potential for tissue motion tracking in 2D-Cine MRI-guided radiotherapy.
  
  \item We propose an approach for modifying and fine-tuning DINOv2 using the LoRA technique, achieving efficient training with minimal additional costs for the 2D-Cine MRI-guided radiotherapy task.
  
  \item Our method, DINOMotion, was evaluated on two different datasets, outperforming other SOTA approaches for tissue motion tracking in 2D-Cine MRI-guided radiotherapy. Notably, DINOMotion demonstrated robustness against large misalignments.
  
\end{itemize}

\section{METHODS AND MATERIALS}

\subsection{DINOMotion}

To achieve accurate motion tracking, our DINOMotion model generates spatial transformations based on matching landmarks that are extracted by the powerful feature representations of DINOv2 \cite{oquab2023DINOv2}, which is efficiently fine-tuned through Low-Rank Adaptation (LoRA) layers. DINOv2 is a foundation model that generates versatile visual features for generalized computer vision tasks. As it was trained on natural images, we adapted DINOv2 to effectively perform landmark-based motion tracking in medical images, ensuring consistent encoding of similar anatomical structures across scans. \textcolor{black}{The primary goal of DINOMotion is to estimate the optimal coordinate transformation \( T_\phi \), a parametric function defined by parameters \( \phi \), that minimizes the discrepancy between the template image (\( \boldsymbol{x}_f \)) and the registered scan (\( \boldsymbol{x}_r = \boldsymbol{x}_m \circ T_{\phi} \)), where \( \boldsymbol{x}_m \) is the moving scan and \( \circ \) denotes spatial transformation. A detailed description of \( T_\phi \) is provided in the \textit{Supplementary Materials}.}

\begin{figure*}[ht]
\centering
\includegraphics[scale=0.37]{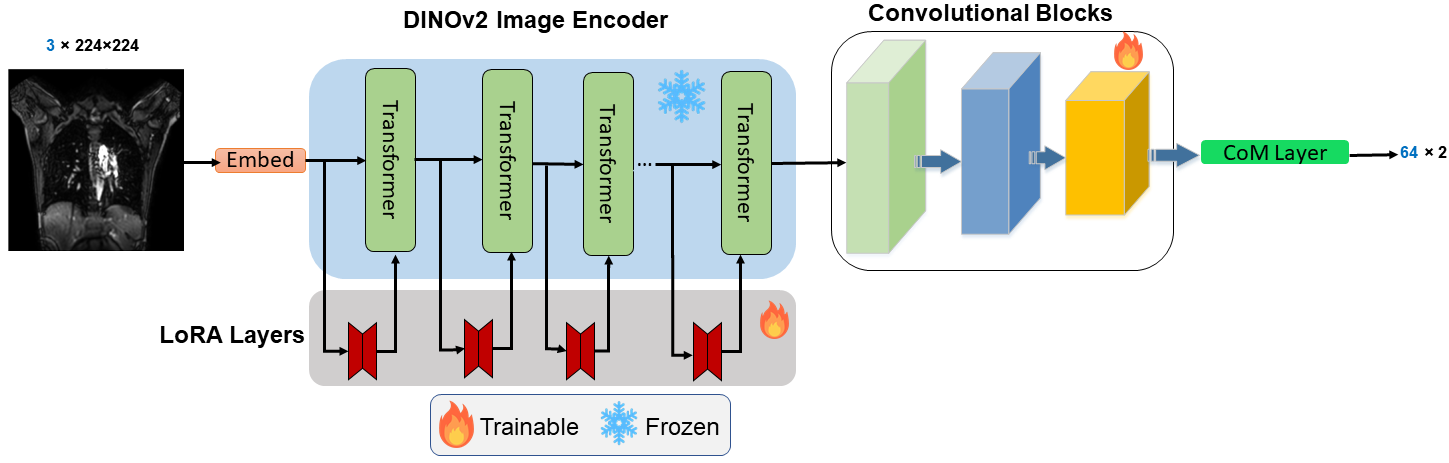}
\caption{The proposed DINOv2-based model for landmark extraction, producing a list of 64 landmarks.}
\label{Network}
\end{figure*}

\textcolor{black}{We adopted the DINOv2 ViT-Base/14 (dinov2\_vitb14) backbone, which consists of 12 Transformer layers with 768-dimensional embeddings, as the core of our landmark extraction model, as illustrated in Fig.~\ref{Network}.} In this model, each scan is first processed through an embedding layer to extract image embeddings, which are then fed into a frozen DINOv2 encoder. \textcolor{black}{To fine-tune the model, we incorporate trainable LoRA layers (rank=4) on top of the frozen DINOv2 encoder. LoRA is a method designed to fine-tune large pre-trained models by introducing trainable low-rank matrices into each layer [16], while keeping the original model weights frozen. This approach enables task-specific adaptation while preserving the generalization capacity of the pre-trained Transformer by constraining updates to a low-rank subspace, thereby minimizing the number of trainable parameters and reducing computational costs. On top of the frozen DINOv2 encoder, we implement a lightweight convolutional decoder consisting of three sequential ConvBlocks. These blocks progressively refine and project the Transformer’s 768-dimensional features down to lower-dimensional spatial representations. Each ConvBlock includes layers with 512, 256, and 64 channels, along with instance normalization, ReLU activation, and max pooling. The use of convolutional layers introduces local spatial context and enhances the precision of landmark localization, a crucial factor in motion tracking where anatomical detail is critical. This architectural split (frozen Transformer backbone with LoRA layers and a trainable convolutional decoder) balances generalization, task-specific learning, and training efficiency.} \textcolor{black}{The network concludes with a center-of-mass (CoM) layer \cite{sofka2017fully,ma2020volumetric} that predicts 64 landmarks, each represented by $(x, y)$ coordinates, resulting in an output of shape $64 \times 2$. This is achieved by computing the center of mass for each activation map in the final feature set (with N =64 channels corresponding to the 64 landmarks). The resulting coordinates are rescaled to the normalized coordinate range of -1 to 1 to match the target resolution during transformation. Subsequently, the predicted landmarks for moving and template scans are being used to estimate a spatial transformation grid via interpolants (thin-plate spline, affine, or rigid). The landmark coordinates themselves are explicitly transformed by applying this estimated transformation. Finally, the landmarks and the moving image are consistently resampled using the same spatial transformation, ensuring alignment consistency.} This design ensures that landmark localization remains consistent with respect to input translations, approximating translation equivariance and enabling precise and robust landmark localization.

\begin{figure*}[ht]
\centering
\includegraphics[scale=0.6]{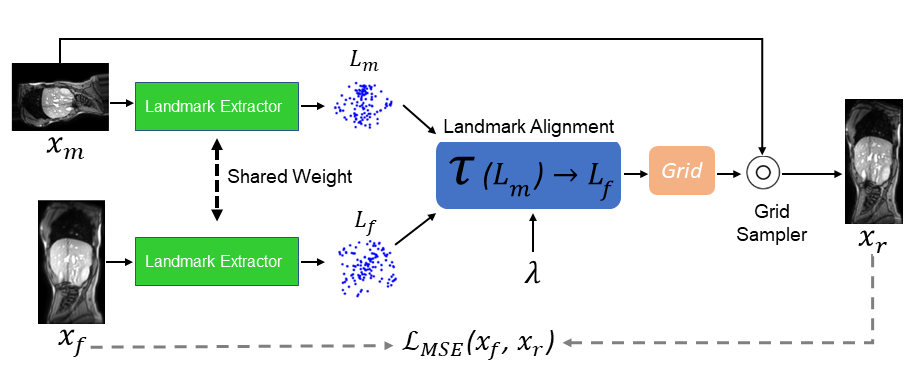}
\caption{The proposed DINOMotion framework for motion tracking. $\boldsymbol{x}_f$, $\boldsymbol{x}_m$, and $\boldsymbol{x}_r$ represent the template, moving, and registered scans, respectively.}
\label{Framework}
\end{figure*}

The full DINOMotion framework is illustrated in Fig. \ref{Framework}, where the template and moving 2D images are passed through a unified landmark detection network (Fig. \ref{Network}), which identifies landmarks in each scan. We offer flexibility in handling both linear and nonlinear transformations. For nonlinear transformation, we have implemented landmark aligners using thin-plate spline (TPS) interpolants with a regularization term ($\lambda$) that controls the smoothness of the deformation. In addition, we also allow affine and rigid interpolants. These landmarks are fed into the respective interpolants to estimate the transformation grid, which is then used to resample the moving scan. Then, the mean squared error (MSE) loss function is applied to compute the error between the template and registered scan.

\subsection{Dataset}

To develop and test our method, we utilized two distinct datasets from sessions of 2D-cine MRI acquisition. The first dataset contains 27 healthy volunteers without radiotherapy interventions, while the second includes 23 patients (data from a previous study \cite{jassar2023real}) who underwent 2D-cine MRI-guided radiotherapy for tumor treatment. For patients, the images were captured across multiple treatment fractions (ranging from 4 to 14 fractions per patient) at various tumor sites, including the liver, kidney, pancreas, prostate, lung, and uterus. The datasets include scans performed under different workflows, such as respiratory (mid-position) and breath-hold (deep inhale/exhale), providing diverse conditions for organ motion and treatment assessment. \textcolor{black}{Both datasets were acquired on a 1.5 T MR‑Linac system (Unity, Elekta AB) using identical real-time orthogonal T1/T2‑weighted cine MRI protocols. The patient dataset was kept separate from training and served solely for external validation.} Informed consent was obtained from all participants.

For each fraction, the first 30 scans from the relevant breathing phase (exhale, mid-position, or static) in both the coronal and sagittal planes were averaged to generate a single template image for each plane. Using different 2D orientations allows better detection of motion in the 3D space. A fixed initial registration was then performed between the pair of 2D images, consisting of the daily extracted 2D MR slice and the 2D cine template image, providing the baseline alignment or translation component. Then, as live 2D cine images are acquired, they can be registered with the template images to estimate tissue motion.

\subsection{Baseline methods}

The Advanced Normalization Tools (ANTs) \cite{avants2009advanced} package is widely regarded as the state-of-the-art (SOTA) in classic medical image registration. In our experiments, we utilized three configurations of ANTs: rigid, affine, and SyN (Symmetric Normalization), the last of which includes an initial rigid and affine alignment step followed by a non-linear transformation. For all ANTs configurations, mutual information (MI) was employed as the similarity metric. Additionally, we benchmarked our approach against NiftyReg \cite{modat2010fast}, which performs non-linear registration following an affine initialization, using normalized mutual information (NMI) as the similarity metric. Lastly, we compared our results with VoxelMorph, a popular deep learning framework optimized for fast and accurate deformable image registration, which leverages CNNs for non-linear registration tasks. For VoxelMorph \cite{balakrishnan2019voxelmorph}, we applied the default parameters specified by the authors and utilized the same training data as in our model.

\subsection{Experimental Setup and Implementation Details}

\textbf{Training:} In our experiments, we arbitrarily selected 17 subjects from the healthy volunteer dataset for training. We obtained a total of 190K pairs of cine MRIs and their corresponding templates from the training subjects across their treatment fractions. All scans were normalized between 0 and 1, resampled to 1 mm resolution, resized to 224 $\times$ 224 pixels, and center-cropped. \textcolor{black}{Since the input MRI images are single-channel, we replicated the grayscale image across all three channels to match the RGB format expected by DINOv2.} We employed the Adam optimizer for model training with a learning rate of $1 \times 10^{-4}$ and a batch size of 32. To optimize the model, we used the MSE loss function to minimize the difference between the template ($\boldsymbol{x}_f$) and the registered scan ($\boldsymbol{x}_r$). Additionally, during training, random affine transformations were applied to the moving images as part of our data augmentation strategy, and $\lambda$ was randomly sampled from a log-uniform distribution ranging between 0 and 10. All learning-based models were implemented in PyTorch and trained on an NVIDIA GeForce RTX 3090 GPU.

\noindent \textbf{Validation:} For validation, we reserved the remaining 10 subjects from the volunteer dataset and the entire second dataset of 23 patients. In the volunteer dataset, we segmented a total of 419 frames, sampled equally across all available fractions, to create segmentations for various organs, including the kidney, liver, pancreas, prostate, and lung, as well as their corresponding template scans. For the patient dataset (comprising 23 patients with 4 kidneys, 8 livers, 9 pancreas, and 2 prostates), we obtained manual segmentations for 224 frames, also sampled equally across all fractions, along with their templates. To ensure high-quality manual segmentations, two expert raters with backgrounds in anatomy and medical imaging (authors CS and LP, referred to as Rater 1 and Rater 2) each performed 50\% of all segmentations. To assess inter-rater variability, they repeated the segmentation of key organs (i.e., the lung, kidney, liver, pancreas, and prostate) on 10 randomly selected frames from all volunteer subjects. The inter-rater variability was assessed by calculating the Dice coefficient between the segmentations performed by Rater 1 and Rater 2 for each scan. The mean Dice scores per organ were as follows: kidney (98\%), liver (95\%), lung (97\%), pancreas (88\%), and prostate (92\%). These high Dice scores demonstrate strong consistency and minimal bias in the segmentation process across different organs, confirming the reliability of the manual segmentation performed by the raters.

\subsection{Evaluation Metrics and Statistical Analysis}
To evaluate the quality of tissue motion tracking of our proposed method, we compared the spatially transformed segmentation by our algorithm to the corresponding template's ground-truth segmentation, using both the Dice Score and Hausdorff Distance as performance metrics. We compared the performance of our method against the baselines using two-sided paired sample t-tests. A p-value of less than 0.05 indicates statistical significance.

\section{Results}

\begin{figure*}[tb]
\centering
\includegraphics[scale=0.37]{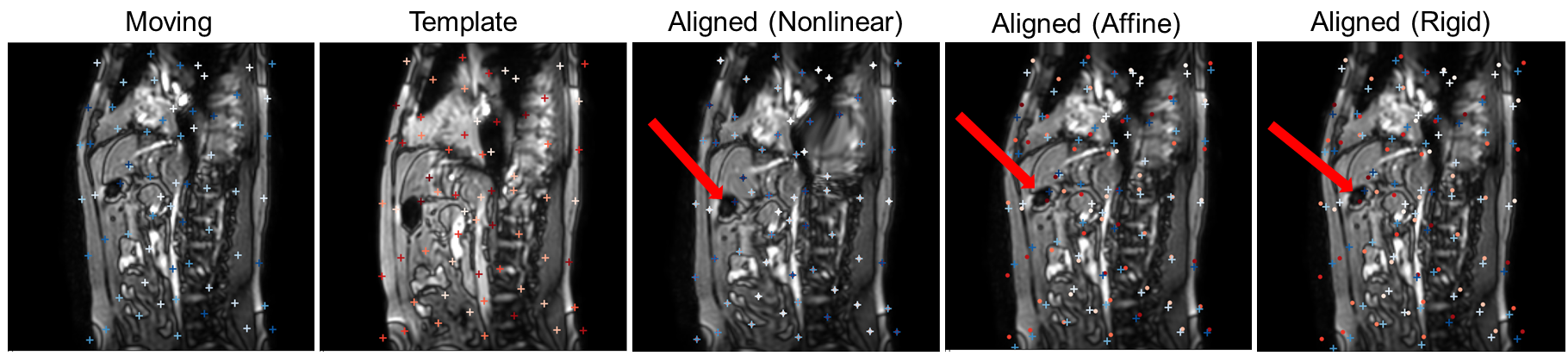}
\caption{Visualization of landmarks for a sample pair of moving and template scans from a volunteer and the aligned scans using DINOMotion under three transformation models: Nonlinear, Affine, and Rigid. Red arrows highlight the differences in motion tracking accuracy among the different transformations.}
\label{SampleOutput}
\end{figure*}

\begin{figure*}[ht]
\centering
\includegraphics[scale=0.165]{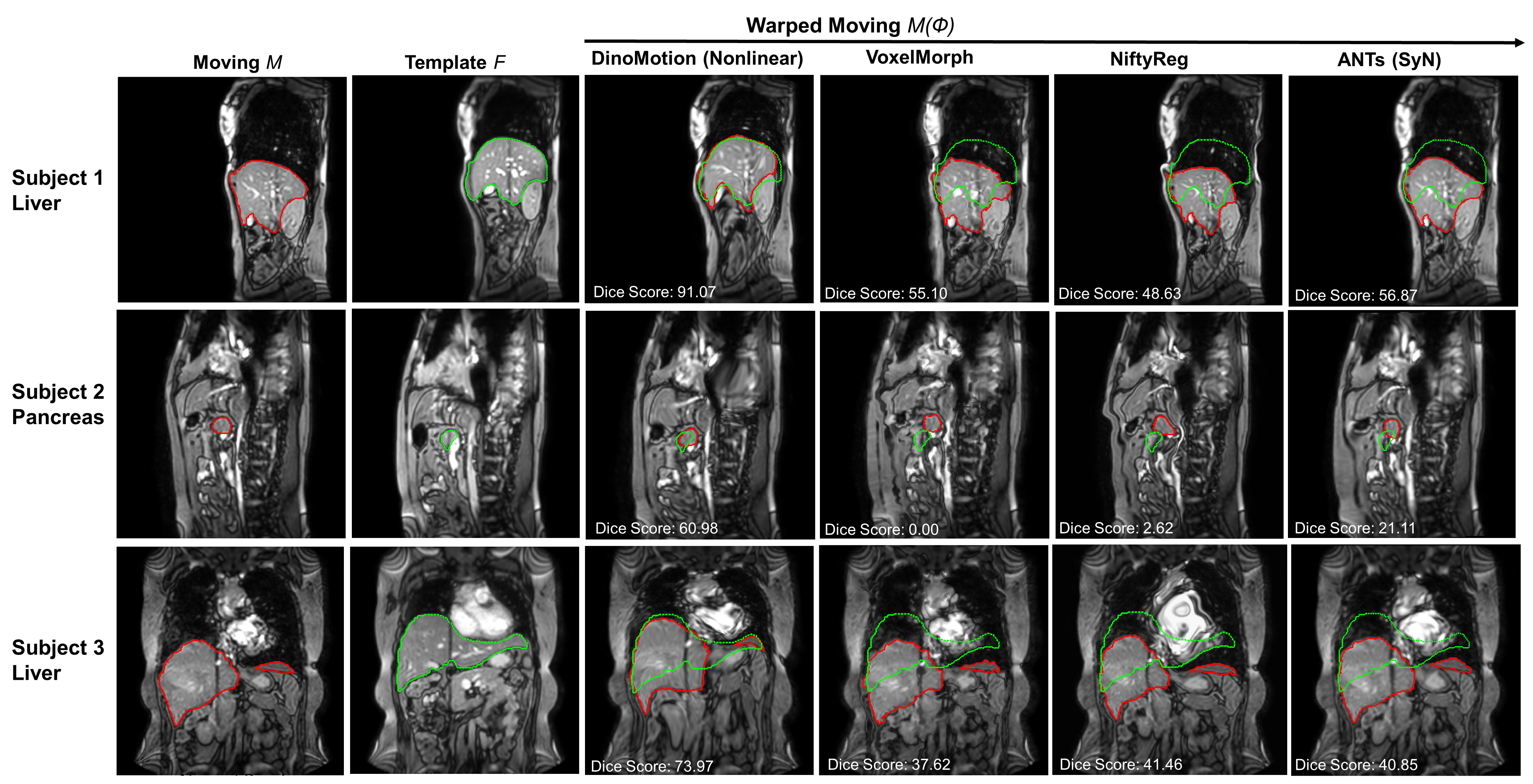}
\caption{\textcolor{black}{Visual comparison of warped scans generated by different methods using samples from the volunteer dataset. Red contours represent the moved segmentations after registration, while green contours indicate the reference (template) segmentations.}}
\label{VisualComp}
\end{figure*}

Table \ref{table1} presents a comparison between the proposed method and baseline approaches in terms of Dice Score and Hausdorff Distance across various organs on the volunteers dataset. NiftyReg achieved reasonable performance, with Dice scores ranging from 50.17\% for the pancreas to 96.19\% for the lung, and Hausdorff distances between 3.88 mm (prostate) and 13.44 mm (pancreas). VoxelMorph showed comparable results but with slightly lower accuracy in some organs. The ANTs framework demonstrated progressive improvements from rigid to affine transformations, with the best results achieved by the SyN variant—particularly notable are the Dice score of 90.20\% and Hausdorff distance of 3.04 mm for the prostate. In contrast, the proposed DINOMotion method outperformed all baselines across most organs. Figure \ref{SampleOutput} illustrates the landmarks detected by DINOMotion for a sample moving and template image pair from a volunteer in the context of tissue motion tracking. The visualization of detected landmarks further highlights the interpretability of the method by providing an intuitive way to assess misalignment and registration accuracy. This qualitative assessment complements quantitative metrics, offering a clearer understanding of where deformations occur and how well the registration aligns anatomical structures across images. The rigid variant achieved high Dice scores, such as 90.67\% for the kidney, and low Hausdorff distances, like 5.88 mm for the kidney. The affine variant showed similar performance. Notably, the nonlinear version of DINOMotion achieved the best results ($p < 0.05$). DINOMotion achieved Dice scores of 92.07\% for the kidney, 90.90\% for the liver, and 95.23\% for the lung. It also recorded the lowest Hausdorff distances in several organs, including 5.47 mm for the kidney and 10.52 mm for the pancreas. \textcolor{black}{Also, Fig. \ref{VisualComp} shows a visual comparison of warped scans generated by different methods using samples from the volunteer dataset, where DINOMotion (Nonlinear) achieves noticeably better motion tracking compared to other approaches. These visual results demonstrate DINOMotion’s robustness in accurately aligning scans despite significant variations in patient breathing patterns encountered in clinical practice.}

\begin{table*}[tb]
  \centering
  \renewcommand{\arraystretch}{2.5}
  
  \resizebox{\textwidth}{!}{
   \Huge 
    \begin{tabular}{|c|c|c|c|c|c|c|c|c|c|c|}
    \hline
    \rowcolor{black} \Huge \textcolor{white}{Method} & \multicolumn{5}{|c|}{\Huge \textcolor{white}{Dice Score (\%) \textcolor{green}{$\uparrow$}}} & \multicolumn{5}{|c|}{\Huge \textcolor{white}{Hausdorff Distance (mm) \textcolor{green}{$\downarrow$}}} \\
    \hline
    \rowcolor{black} \Huge \textcolor{white}{Organ} & \Huge \textcolor{white}{Kidney} & \Huge \textcolor{white}{Liver} & \Huge \textcolor{white}{Lung} & \Huge \textcolor{white}{Pancreas} & \Huge \textcolor{white}{Prostate} & \Huge \textcolor{white}{Kidney} & \Huge \textcolor{white}{Liver} & \Huge \textcolor{white}{Lung} & \Huge \textcolor{white}{Pancreas} & \Huge \textcolor{white}{Prostate} \\
    \hline
    \Huge NiftyReg & \Huge 86.38 $\pm$ 17.12 & \Huge 88.66 $\pm$ 13.94 & \Huge \textbf{96.19 $\pm$ 2.94} & \Huge 50.17 $\pm$ 36.94 & \Huge 87.10 $\pm$ 9.38 & \Huge 8.15 $\pm$ 8.04 & \Huge 9.45 $\pm$ 7.76 & \Huge 6.86 $\pm$ 5.23 & \Huge 13.44 $\pm$ 11.40 & \Huge 3.88 $\pm$ 2.34 \\
    
    \Huge VoxelMorph & \Huge 85.81 $\pm$ 16.36 & \Huge 84.95 $\pm$ 16.10 & \Huge 94.46 $\pm$ 3.64 & \Huge 42.21 $\pm$ 35.27 & \Huge 87.98 $\pm$ 5.96 & \Huge 8.68 $\pm$ 7.85 & \Huge 10.46 $\pm$ 8.51 & \Huge 8.37 $\pm$ 5.04 & \Huge 14.57 $\pm$ 11.35 & \Huge 3.54 $\pm$ 1.53 \\
    
    \Huge ANTS (Rigid) & \Huge 83.70 $\pm$ 16.94 & \Huge 85.16 $\pm$ 14.47 & \Huge 93.41 $\pm$ 3.54 & \Huge 40.23 $\pm$ 32.45 & \Huge 82.51 $\pm$ 7.59 & \Huge 5.52 $\pm$ 5.45 & \Huge 9.81 $\pm$ 7.71 & \Huge 8.12 $\pm$ 4.62 & \Huge 15.40 $\pm$ 11.14 & \Huge 4.34 $\pm$ 1.22 \\
    
    \Huge ANTS (Affine) & \Huge 83.78 $\pm$ 16.74 & \Huge 85.26 $\pm$ 14.90 & \Huge 93.90 $\pm$ 3.33 & \Huge 47.45 $\pm$ 32.68 & \Huge 82.88 $\pm$ 7.65 & \Huge 5.52 $\pm$ 5.20 & \Huge 9.62 $\pm$ 7.79 & \Huge 7.84 $\pm$ 4.49 & \Huge 13.76 $\pm$ 10.34 & \Huge 4.30 $\pm$ 1.20 \\
    
    \Huge ANTS (SyN) & \Huge 86.26 $\pm$ 16.15 & \Huge 88.94 $\pm$ 14.04 & \Huge 95.94 $\pm$ 2.28 & \Huge 54.24 $\pm$ 34.74 & \Huge \textbf{90.20 $\pm$ 3.48} & \Huge 5.49 $\pm$ 6.23 & \Huge 8.69 $\pm$ 7.93 & \Huge \textbf{6.30 $\pm$ 4.17} & \Huge 12.56 $\pm$ 10.79 & \Huge \textbf{3.04 $\pm$ 0.94} \\
    
    \Huge \textbf{DINOMotion (Rigid)} & \Huge 90.67 $\pm$ 6.74 & \Huge 89.03 $\pm$ 10.87 & \Huge 93.81 $\pm$ 3.59 & \Huge 52.49 $\pm$ 30.12 & \Huge 83.37 $\pm$ 8.92 & \Huge 5.88 $\pm$ 3.99 & \Huge \textbf{7.81 $\pm$ 6.43} & \Huge 7.14 $\pm$ 3.87 & \Huge 11.67 $\pm$ 9.90 & \Huge 3.59 $\pm$ 1.34 \\
    
    \Huge \textbf{DINOMotion (Affine)} & \Huge 90.38 $\pm$ 7.01 & \Huge 89.19 $\pm$ 10.56 & \Huge 94.63 $\pm$ 3.15 & \Huge 53.64 $\pm$ 29.96 & \Huge 83.19 $\pm$ 7.36 & \Huge 5.98 $\pm$ 4.09 & \Huge 8.48 $\pm$ 7.55 & \Huge 6.93 $\pm$ 3.91 & \Huge 11.49 $\pm$ 9.88 & \Huge 3.58 $\pm$ 1.18 \\
    
    \Huge \textbf{DINOMotion (Nonlinear)} & \Huge \textbf{92.07 $\pm$ 5.62}\bigasterisk & \Huge \textbf{90.90 $\pm$ 7.22}\bigasterisk & \Huge 95.23 $\pm$ 3.05 & \Huge \textbf{60.04 $\pm$ 29.03}\bigasterisk & \Huge 82.76 $\pm$ 7.47 & \Huge \textbf{5.47 $\pm$ 3.54}\bigasterisk & \Huge 8.31 $\pm$ 7.00 & \Huge 6.72 $\pm$ 3.83 & \Huge \textbf{10.52 $\pm$ 9.42}\bigasterisk & \Huge 3.76 $\pm$ 1.11 \\
    
    \hline
    \end{tabular}
  }
  \caption{Quantitative results for Dice Score and Hausdorff Distance (mean$\pm$std) across various organs on the volunteer's dataset. $*$ indicates statistically significant differences (p-value < 0.05).}
  \label{table1}
\end{table*}

\noindent
Table \ref{TableTime} compares the inference times of the proposed method with baseline methods on both CPU and GPU. DINOMotion demonstrated competitive computational times, with GPU execution times ranging from 0.028 to 0.030 seconds, which is faster than NiftyReg's 0.26 seconds and only slightly slower than VoxelMorph's 0.004 seconds. On the CPU, DINOMotion required 0.35 to 0.39 seconds, which is faster than ANTs (SyN) at 2.17 seconds and NiftyReg at 0.67. Overall, DINOMotion demonstrated superior performance, particularly in its nonlinear configuration, by achieving higher accuracy in tissue alignment without significant computational overhead.

\begin{table}[tb]
\centering
\caption{Comparison of the proposed method with baseline methods regarding mean inference time on GPU and CPU on the volunteers dataset.}

\begin{tabular}{|c|c|c|}
\rowcolor{black} 
\hline 
\color{white} Method & \color{white} CPU Time (s) \textcolor{green}{$\downarrow$} & \color{white} GPU Time (s) \textcolor{green}{$\downarrow$}\\
\hline 
NiftyReg & 0.67 & 0.26 \\
\hline 
VoxelMorph & \textbf{0.058} & \textbf{0.004} \\
\hline 
ANTS (Rigid) & 0.74 & -\\
\hline 
ANTS (Affine) & 0.77 & -\\
\hline 
ANTS (SyN) & 2.17 & -\\
\hline 
\textbf{DINOMotion (Rigid)} & 0.35 & 0.028 \\
\hline 
\textbf{DINOMotion (Affine)} & 0.35 & 0.029 \\
\hline 
\textbf{DINOMotion (Nonlinear)} & 0.39 & 0.030 \\
\hline 
\end{tabular}

\label{TableTime}
\end{table}

\begin{figure*}[tb]
\centering
\includegraphics[scale=0.35]{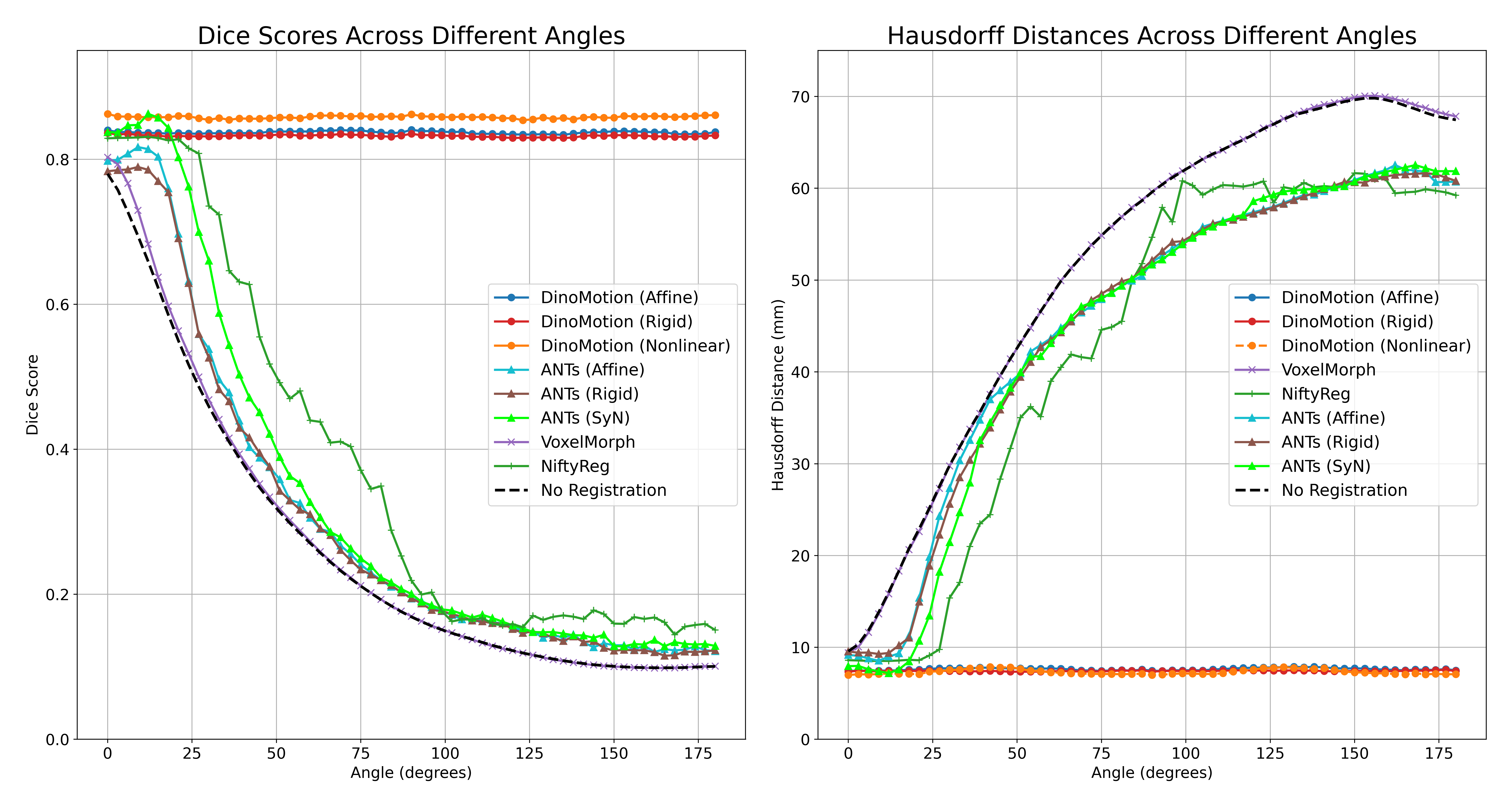}
\caption{Comparison of mean Dice score and Hausdorff distance across different initial rotation misalignments of the moving image for baseline models and all DINOMotion variants over the volunteer dataset.}
\label{ComparisonPlot}
\end{figure*}

\begin{figure*}[tb]
\centering
\includegraphics[scale=0.35]{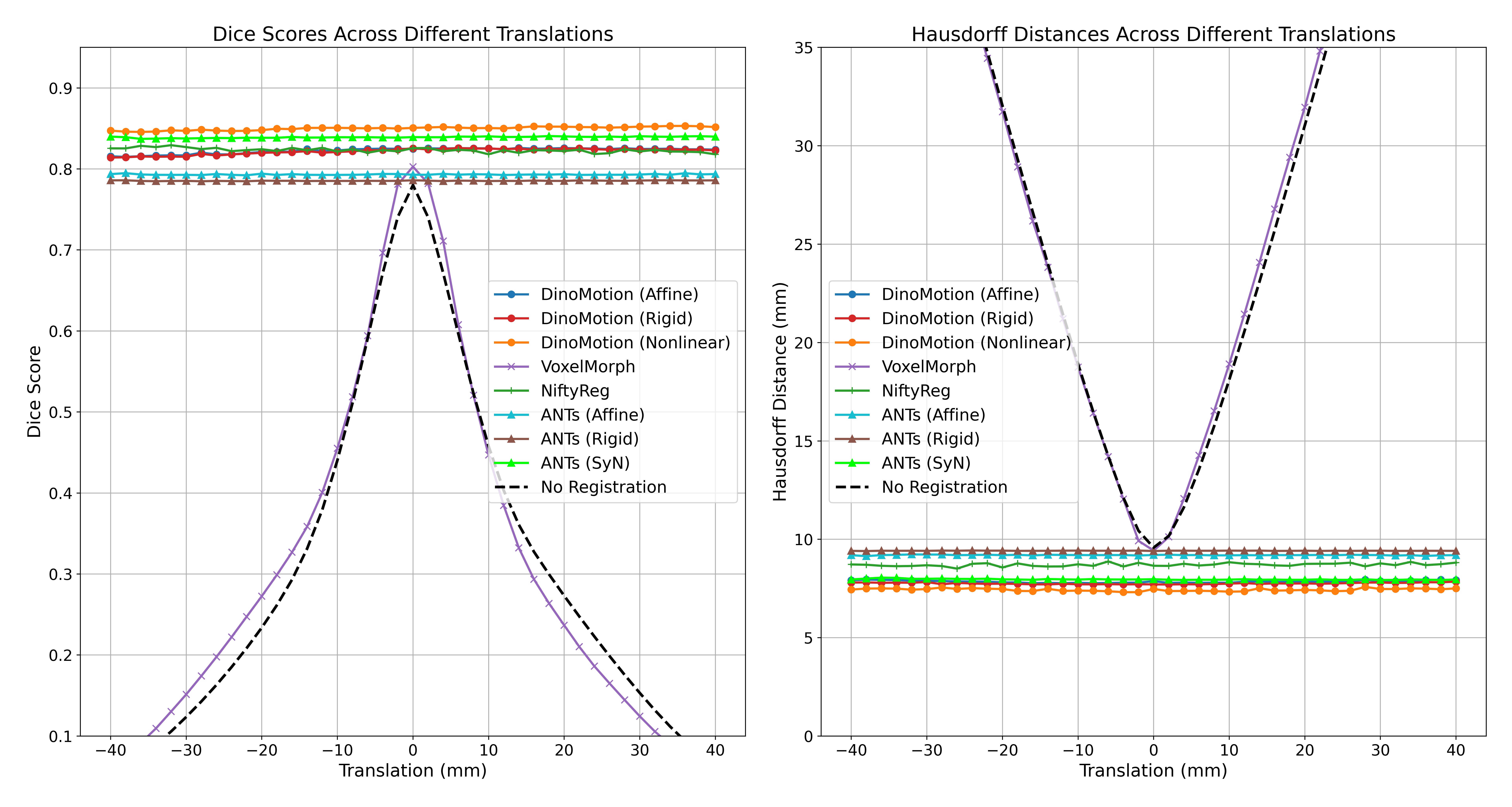}
\caption{\textcolor{black}{Comparison of mean Dice score and Hausdorff distance across different initial translational misalignments of the moving image for baseline models and all DINOMotion variants over the volunteer dataset.}}
\label{ComparisonPlot2}
\end{figure*}

To assess the robustness of DINOMotion, we examine its performance, along with that of the baseline models, under conditions of augmenting significant misalignment to the existing moving scans, particularly focusing on rotation and \textcolor{black}{translation}, important factors in MRI-guided radiotherapy where large subject motions can challenge tracking algorithms. \textcolor{black}{Figures \ref{ComparisonPlot} and \ref{ComparisonPlot2} illustrate the mean Dice score and Hausdorff distance as functions of the augmented rotation and translation misalignment for both the baselines and all DINOMotion variants.} Our analysis reveals that while models, such as NiftyReg, VoxelMorph, and all ANTs variants, perform adequately with minimal augmented rotation (e.g., around 0 degrees of rotation), their performance declines significantly as misalignment increases. In contrast, all DINOMotion variants demonstrated strong and consistent performance across a wide range of transformations, even with large \textcolor{black}{rotation} misalignments, without notable performance decline. \textcolor{black}{Additionally, while VoxelMorph remains stable under small translational shifts (e.g., around 0 mm), its performance degrades sharply as translational misalignment increases. However, DINOMotion (Nonlinear) consistently achieves the highest Dice scores and the lowest Hausdorff distances and outperforms all comparator models even under severe translations.}

To validate the effectiveness of DINOMotion on tumor cases, we evaluated its performance across all transformation configurations (Rigid, Affine, and Nonlinear) using the patient dataset. Table \ref{table2} provides a registration accuracy comparison between the proposed method and baseline approaches. As demonstrated in Table \ref{table2}, the nonlinear configuration of DINOMotion achieved the highest Dice Scores, with 87.04\% $\pm$ 11.54\% for the pancreas and 92.71\% $\pm$ 8.51\% for the liver, and the lowest Hausdorff Distances, 4.55 $\pm$ 4.14 for the prostate and 5.75 $\pm$ 4.55 for the pancreas. 
DINOMotion outperformed other methods, including ANTS (SyN), VoxelMorph, and NiftyReg ($p < 0.05$) in nonlinear registration.  Even in its Rigid and Affine configurations, DINOMotion demonstrated competitive performance, consistently matching or surpassing the baseline methods in both metrics. For example, DINOMotion (Affine) achieved a Dice Score of 83.16\% $\pm$ 15.69\% for the kidney and 88.82\% $\pm$ 10.31\% for the liver, with a Hausdorff Distance of 6.70 $\pm$ 4.55 for the pancreas. These results highlight the robustness of DINOMotion across different transformation models. Overall, this comprehensive evaluation underscores DINOMotion’s potential for enhancing motion tracking in clinical settings, particularly in radiotherapy applications, where precise alignment of anatomical structures is crucial for treatment outcomes. \textcolor{black}{The strong performance on the patient cohort highlights the DinoMotion robustness and generalizability in real clinical use.}

\begin{table*}[tb]
  \centering
  \renewcommand{\arraystretch}{1.5}
  
  \resizebox{\textwidth}{!}{
   \Huge 
    \begin{tabular}{|c|c|c|c|c|c|c|c|c|}
    \hline
    \rowcolor{black} \Huge \textcolor{white}{Method} & \multicolumn{4}{|c|}{\Huge \textcolor{white}{Dice Score (\%) \textcolor{green}{$\uparrow$}}} & \multicolumn{4}{|c|}{\Huge \textcolor{white}{Hausdorff Distance (mm) \textcolor{green}{$\downarrow$}}} \\
    \hline
    \rowcolor{black} \Huge \textcolor{white}{Organ} & \Huge \textcolor{white}{Kidney} & \Huge \textcolor{white}{Liver} & \Huge \textcolor{white}{Pancreas} & \Huge \textcolor{white}{Prostate} & \Huge \textcolor{white}{Kidney} & \Huge \textcolor{white}{Liver} & \Huge \textcolor{white}{Pancreas} & \Huge \textcolor{white}{Prostate} \\
    \hline
    \Huge NiftyReg & \Huge 80.02 $\pm$ 15.39  & \Huge 87.00 $\pm$ 12.26  & \Huge 79.05 $\pm$ 14.78  & \Huge 46.70 $\pm$ 29.13  & \Huge 7.82 $\pm$ 7.45  & \Huge 8.03 $\pm$ 6.10  & \Huge 8.92 $\pm$ 15.90  & \Huge 7.87 $\pm$ 7.27 \\
    
    \Huge VoxelMorph & \Huge 82.28 $\pm$ 15.15 & \Huge 90.37 $\pm$ 8.61 & \Huge 80.98 $\pm$ 12.74 & \Huge 67.12 $\pm$ 28.31 & \Huge 7.72 $\pm$ 7.24 & \Huge 7.44 $\pm$ 5.42 & \Huge 7.60 $\pm$ 5.10 & \Huge 5.24 $\pm$ 3.65 \\
    
    \Huge ANTS (Rigid) & \Huge 80.45 $\pm$ 15.78   & \Huge 87.84 $\pm$ 10.48  & \Huge 78.72 $\pm$ 15.68  & \Huge 45.55 $\pm$ 30.07  & \Huge 7.42 $\pm$ 6.49  & \Huge 7.86 $\pm$ 5.82  & \Huge 7.41 $\pm$ 5.30  & \Huge 6.78 $\pm$ 4.33 \\

    \Huge ANTS (Affine) & \Huge 82.65 $\pm$ 14.12   & \Huge 89.60 $\pm$ 7.41  & \Huge 79.75 $\pm$ 15.43  & \Huge 48.47 $\pm$ 30.70  & \Huge 7.00 $\pm$ 5.68  & \Huge 6.82 $\pm$ 4.70  & \Huge 6.72 $\pm$ 4.65  & \Huge 6.25 $\pm$ 4.67 \\
    
    \Huge ANTS (SyN) & \Huge \textbf{85.83 $\pm$ 14.11} & \Huge 92.56 $\pm$ 8.50 & \Huge 86.12 $\pm$ 11.47 & \Huge 66.27 $\pm$ 25.17 & \Huge 6.81 $\pm$ 6.95 & \Huge 6.54 $\pm$ 5.81 & \Huge 6.11 $\pm$ 4.19 & \Huge 5.59 $\pm$ 3.16\\
    
    \Huge \textbf{DINOMotion (Rigid)} & \Huge 82.96 $\pm$ 15.78 & \Huge 88.24 $\pm$ 10.81 & \Huge 80.16 $\pm$ 14.04 & \Huge 44.35 $\pm$ 32.41 & \Huge 7.50 $\pm$ 7.10 & \Huge 7.42 $\pm$ 5.72 & \Huge 6.96 $\pm$ 4.49 & \Huge 6.14 $\pm$ 3.59  \\
    
    \Huge \textbf{DINOMotion (Affine)} & \Huge 83.16 $\pm$ 15.69 & \Huge 88.82 $\pm$ 10.31 & \Huge 81.76 $\pm$ 13.42 & \Huge 45.96 $\pm$ 31.57 & \Huge 7.39 $\pm$ 7.04 & \Huge 7.27 $\pm$ 5.66 & \Huge 6.70 $\pm$ 4.55 & \Huge 6.01 $\pm$ 3.59  \\
    
    \Huge \textbf{DINOMotion (Nonlinear)} & \Huge 84.34 $\pm$ 14.94 & \Huge \textbf{92.71 $\pm$ 8.51}\bigasterisk & \Huge \textbf{87.04 $\pm$ 11.54}\bigasterisk & \Huge \textbf{72.03 $\pm$ 27.66}\bigasterisk & \Huge \textbf{6.78 $\pm$ 7.06}\bigasterisk & \Huge \textbf{6.18 $\pm$ 5.62}\bigasterisk & \Huge \textbf{5.75 $\pm$ 4.55}\bigasterisk & \Huge \textbf{4.55 $\pm$ 4.14}\bigasterisk  \\
    
    \hline
    \end{tabular}
  }
  \caption{Quantitative results for Dice Score and Hausdorff Distance (mean$\pm$std) across various organs on the patient dataset. $*$ indicates statistically significant differences (p-value < 0.05).}
  \label{table2}
\end{table*}

\section{Discussion}

\subsection{Strengths of DINOMotion}

The results of this study demonstrate the potential of DINOMotion in advancing tissue motion tracking for 2D-Cine MRI-guided radiotherapy. By utilizing the DINOv2 model, a powerful self-supervised vision framework, DINOMotion effectively extracts landmarks for accurate motion tracking. Compared to established methods such as ANTS, NiftyReg, and VoxelMorph, DINOMotion shows superior performance, particularly in scenarios involving large misalignments—one of the most critical challenges in MRI-guided radiotherapy. 

A key strength of DINOMotion lies in its ability to handle both linear and nonlinear transformations directly, enabling accurate estimation of tissue motion. Our findings show that DINOMotion's nonlinear configuration consistently outperformed baseline methods on both the volunteer and patient datasets with a much lower computational load. Specifically, DINOMotion achieved Dice scores exceeding 90\% for organs like the liver and kidney and reduced the Hausdorff distance by up to 20\% compared to baseline methods. This is especially important for real-time clinical applications, where tissue deformations are often complex and nonlinear, and precise tracking is essential for effective dose delivery.

One of the significant advantages of DINOMotion is its capability to handle large movements, which are common in abdominal and thoracic imaging due to patient breathing and organ motion. Unlike traditional deep learning models and classical registration methods, such as ANTS and NiftyReg, which often struggle with large displacements, DINOMotion demonstrated consistent performance across a wide range of initial rotation \textcolor{black}{and translation} misalignments. This robustness is particularly critical in radiotherapy, where accurately detecting deep respiration is essential for pausing the treatment beam and preventing unintended dose delivery to surrounding organs. By reliably tracking large respiratory-induced shifts, DINOMotion enhances motion management in dynamic clinical settings.

\subsection{Clinical Relevance}

\textcolor{black}{In MRI-guided radiotherapy and related clinical practice, a Dice score greater than 85\% is generally considered sufficient for proper organ localization and motion tracking in abdominal and pelvic treatments \cite{bordigoni2024automated, zhang2020patient, mackay2023review}. This threshold helps ensure accurate targeting while minimizing radiation exposure to surrounding healthy tissues. While a Dice score of 85\% may be sufficient in some settings, improvements beyond this threshold (such as our reported increase from 86.38\% to 92.07\% for kidney tracking) can still make a meaningful difference depending on the location and extent of the tumor. Higher accuracy reduces variability across fractions, improves the precision of adaptive planning, and allows for potential reduction in treatment margins, thereby better sparing adjacent organs-at-risk. It is also important to note that baseline methods often fail to achieve this clinically sufficient threshold when the deformation to estimate is large, such as during deep inspiration. Another example of large deformation arises from bladder filling in pelvic treatments. This is particularly important in workflows requiring daily adaptation or tight margin radiotherapy.}

\textcolor{black}{The clinical relevance of registration accuracy varies across organs and treatment contexts. For abdominal radiotherapy, high accuracy for the liver, pancreas, and kidney is especially critical, as these are often primary targets or nearby organs-at-risk. In contrast, the lung typically appeared peripherally (only partially visible in the field-of-view of liver/pancreas images) and was not the focus of registration, which explains why DINOMotion's performance was slightly lower for the lung compared to these abdominal organs. Despite this, DINOMotion still achieved strong lung performance (e.g., Dice score of 95.23\%) while offering robustness to large misalignments and improved interpretability.}

\textcolor{black}{Figure. \ref{VisualComp} shows DINOMotion’s (Nonlinear) ability to handle challenging motion scenarios, including both in-plane and through-plane motions. In particular, Subject 2 demonstrates a particularly challenging case involving both in-plane and through-plane motion, while Subject 3 highlights difficulties arising from through-plane motion. Despite these complexities, DINOMotion (Nonlinear) maintains robust alignment and delivers accurate results while competing methods fail, resulting in higher Dice scores. These qualitative results illustrate that the observed performance gains translate to clinically meaningful improvements in localization precision, especially for smaller or more deformable organs, such as the pancreas. Moreover, accurately handling these realistic clinical scenarios ensures proper dose delivery, significantly reducing the risk of unintended irradiation of adjacent healthy tissues.}

\subsection{Interpretability}

DINOMotion’s interpretability sets it apart from many black-box deep learning models. By detecting landmarks and computing transformations based on these points, the model provides greater explainability of the motion-tracking technique, making it more suitable for clinical use. \textcolor{black}{In contrast to post-hoc interpretability techniques like saliency maps \cite{simonyan2013deep} or Grad-CAM \cite{selvaraju2017grad,selvaraju2020grad}, which offer indirect and often ambiguous explanations, DINOMotion provides intrinsic interpretability through its explicit landmark pair predictions, enabling direct visualization of anatomical correspondences.} This interpretability not only enhances clinician trust but also facilitates easier debugging and refinement of the model in clinical workflows, which is crucial in risk-sensitive medical contexts.

\subsection{Limitations \& Future Directions}

\textcolor{black}{While DINOMotion currently has a higher inference time than VoxelMorph on GPU ($\sim$ 30ms vs. $\sim$ 4 ms), it is still well within the real-time requirements for clinical workflows in 2D-Cine MRI-guided radiotherapy. Notably, DINOMotion offers a substantial improvement in accuracy over VoxelMorph for several key organs, which could justify its higher runtime for the precision-driven clinical setting. Furthermore, DINOMotion is considerably faster than other classical methods (e.g., ANTs-SyN at $\sim$ 2.17s on CPU), which, despite their high computational cost, are still often used in clinical settings. In our future work, we plan to explore model distillation \cite{fundel2025distillation} or pruning \cite{eliopoulos2025pruning} to compress the DINOv2 backbone without sacrificing accuracy. Nevertheless, with its promising results, DINOMotion still has a few limitations. First, the current model was trained and validated using data from a single institution and scanner due to availability constraints, which limit its ability to generalize across inter-scanner or multi-center settings.} \textcolor{black}{To address this limitation, future studies will incorporate diverse multi-institutional datasets to enhance generalizability. Second, the current framework is designed for individual 2D slices at a single time point and does not directly leverage temporal or volumetric contexts, which could be instrumental for handling complex deformations across time or in 3D (e.g., through-plane motion). Future improvements include integrating temporal and volumetric (3D) contexts to better model complex deformations. Specifically, DINOMotion’s landmark-based design is well-suited for workflows such as Anatomic Position Monitoring (APM) \cite{jassar2023real} and True Tracking, where 3D MRI volumes are acquired daily and compared against reference templates. By adapting the current architecture with 3D convolutional decoders and extending the center-of-mass prediction to volumetric space, the model could potentially facilitate interpretable 2D–3D or full 3D registration. Besides, leveraging landmarks predicted in both volumetric and template data could enable robust and clinically interpretable registration, potentially enhancing the accuracy and reliability of motion management strategies. Third, although we fine-tune the DINOv2 backbone for medical landmark detection, the overall dataset size remains relatively small, which reduces robustness, particularly in cases of irregular motion or limited structural correspondence between template and moving scans. Future work will scale fine-tuning efforts to larger and more diverse medical imaging datasets to improve accuracy and robustness. Finally, predicted landmarks are currently not organ-specific and are used with equal importance in deformation estimation. Consequently, common failure scenarios include scans with rapid or irregular motion, where partial structural correspondence between moving and template scans exists. Future studies could adopt organ-specific landmark prediction strategies to improve localization accuracy, especially for anatomically complex organs. Additionally, incorporating uncertainty estimation techniques could further enhance robustness in regions with limited correspondence or ambiguous anatomical boundaries. Lastly, while DINOMotion demonstrates strong overall performance, we observe higher standard deviations in Dice and Hausdorff distance for certain organs, as shown in Table I. These variances largely reflect inter-subject and intra-organ differences in anatomy and motion characteristics across the dataset. In clinical contexts, such variability is especially pronounced for anatomically complex or low-contrast structures like the pancreas and prostate, which pose inherent challenges for consistent registration. Importantly, standard deviation offers insight into the reliability of a model’s predictions across patients. This variability can be primarily attributed to the limited dataset size, irregular motion patterns (through-plane motions, etc), and possible incomplete structural correspondence.}

\section{Conclusions}
We introduced DINOMotion, a novel DL-driven approach for motion tracking in 2D-Cine MRI-guided radiotherapy, which leverages corresponding landmarks to compute the optimal transformation for aligning live cine scans to a template. Additionally, DINOMotion shows strong robustness to significant misalignments and enhanced interpretability. We validated these strengths through experiments on real-world datasets of 2D-Cine MRI-guided radiotherapy scans against SOTA baselines. DINOMotion's superior performance and beneficial strengths make it a valuable tool for improving the safety and outcomes of MR-guided radiotherapy treatments.

\section{Acknowledgments }
This work was supported by Elekta AB and Mitacs. Also, all procedures performed in studies involving human participants were in accordance with the ethical standards of the University Human Research Ethics Committee, and informed consent was obtained from every participant.


\bibliographystyle{IEEEtran}
\bibliography{ref.bib}



\end{document}


\title{DINOMotion: advanced robust tissue motion tracking with DINOv2 in 2D-Cine MRI-guided radiotherapy\\
Supplementary Material}
\author{Soorena Salari, Catherine Spino, Laurie-Anne Pharand, Fabienne Lathuiliere, Hassan Rivaz \IEEEmembership{Senior Member, IEEE}, Silvain Beriault, Yiming Xiao \IEEEmembership{Senior Member, IEEE}
\thanks{Soorena Salari and Yiming Xiao are with the Department of Computer Science and Software Engineering, Concordia University, Montreal, Canada (e-mail: soorena.salari@concordia.ca and yiming.xiao@concordia.ca).}
\thanks{Hassan Rivaz is with the Department of Electrical and Computer Engineering, Concordia
University, Montreal, Canada (e-mail: hassan.rivaz@concordia.ca).}
\thanks{Catherine Spino, Laurie-Anne Pharand, and Fabienne Lathuiliere were with the Elekta Ltd., Montreal, Canada (e-mail: catherine.spino@polymtl.ca, laurie-anne.pharand@polymtl.ca, and fabienne.lathuiliere@gmail.com).}
\thanks{Silvain Beriault was with Elekta Ltd., Montreal, Canada. He is now with Diagnos Medical Systems, Montreal, QC, Canada (e-mail: silvain.beriault@gmail.com).}
}

\maketitle
\section{METHODS AND MATERIALS}

\subsection{Differentiable Coordinate Transformations via Analytical Methods}

\textbf{Notation:} 
Lowercase and uppercase bold symbols represent column vectors and matrices, respectively. A landmark in \(D\)-dimensional space is represented as \(\mathbf{X}_{mov} \in \mathbb{R}^D\). Its homogeneous coordinate form is given by \(\tilde{\mathbf{X}}_{mov} = [\mathbf{X}_{mov},\, 1]^T\). Superscripts, such as \(\mathbf{X}_{mov}^{(i)}\), refer to distinct instances (e.g., samples), while subscripts, like \(\mathbf{X}_{mov,i}\), indicate the \(i\)-th component of the vector.

We consider a family of parametric transformations that can be derived explicitly from landmark correspondences. Suppose there are \(M\) matching landmark pairs \(\{(\mathbf{X}_{mov}^{(i)}, \mathbf{X}_{fixed}^{(i)})\}_{i=1}^{M}\), where \(\mathbf{X}_{mov}^{(i)},\, \mathbf{X}_{fixed}^{(i)} \in \mathbb{R}^D\) and \(M > D\). For convenience, define 
$ 
\mathbf{X}_{mov}^{all} := \langle \mathbf{X}_{mov}^{(1)}, \ldots, \mathbf{X}_{mov}^{(M)} \rangle \in \mathbb{R}^{D \times M},
$
and similarly, let \(\tilde{\mathbf{X}}_{mov}^{all}\) and \({\mathbf{X}}_{fixed}^{all}\) denote the homogeneous moving scan and fixed scan landmark sets, respectively. We then introduce a transformation function 
$ 
T_\varphi : \mathbb{R}^D \rightarrow \mathbb{R}^D,
$
with transformation parameters \(\varphi\). 

\subsubsection{Rigid}

Rigid transformations consist of a rotation and a translation. Let 
$\mathbf{R} \in \mathbb{R}^{D \times D}$ be a rotation matrix and  $ \mathbf{t} \in \mathbb{R}^{D \times 1} $ a translation vector. The rigid mapping is then defined as
\begin{equation}
    T_\varphi(\mathbf{X}_{mov}) = \mathbf{R}\,\mathbf{X}_{mov} + \mathbf{t},
\end{equation}
with the parameter set $\varphi = \{\mathbf{R}, \mathbf{t}\}.$

The optimal translation can be determined by subtracting the centroids of the moving and fixed landmark sets. The centroids can be calculated by averaging the moving and fixed landmark pairs:
$ 
\bar{\mathbf{X}}_{mov} = \frac{1}{M} \sum_{i=1}^{M} \mathbf{X}_{mov}^{(i)} \quad \text{and} \quad \bar{\mathbf{X}}_{fixed} = \frac{1}{M} \sum_{i=1}^{M} \mathbf{X}_{fixed}^{(i)},
$
so that the optimal translation is $ \mathbf{t}^* = \bar{\mathbf{X}}_{fixed} - \bar{\mathbf{X}}_{mov} .$
    
The task of determining the most accurate rotational alignment between two sets of points has been extensively investigated and is commonly referred to as the orthogonal Procrustes problem \cite{viklands2006algorithms}. To compute this rotation, the first step involves centering the landmark sets by subtracting their respective centroids
$
\tilde{\mathbf{X}}_{mov}^{(i)} = \mathbf{X}_{mov}^{(i)} - \bar{\mathbf{X}}_{mov} \quad \text{and} \quad \tilde{\mathbf{X}}_{fixed}^{(i)} = \mathbf{X}_{fixed}^{(i)} - \bar{\mathbf{X}}_{fixed}.
$
Secondly, cross-correlation matrix is obtained: $ \mathbf{\Sigma} = \tilde{\mathbf{X}}_{mov}^{all}\, \tilde{\mathbf{X}}_{fixed}^{all\,T},$. Next, we can compute the singular value decomposition (SVD) of \(\mathbf{\Sigma}\) as
$ \mathbf{\Sigma} = \mathbf{U}\,\mathbf{\Lambda}\,\mathbf{V}^T $. Finally, the optimal rotation is obtained by
\begin{equation}
    \mathbf{R}^* = \mathbf{V}\,\mathbf{U}^T.
\end{equation}


\subsubsection{Affine}

Affine transformations can be described using matrix multiplication. Let $ \mathbf{B} \in \mathbb{R}^{D \times (D+1)} $ be the affine transformation matrix. Then, given a moving landmark expressed in homogeneous coordinates, the affine mapping is defined as

\begin{equation}
    T_\varphi(\mathbf{X}_{mov}) = \mathbf{B}^T \tilde{\mathbf{X}}_{mov},
\end{equation}
where the parameter set is $\varphi = \{\mathbf{B}\}.$

Given \(M\) corresponding landmark pairs 
$ \{(\mathbf{X}_{mov}^{(i)}, \mathbf{X}_{fixed}^{(i)})\}_{i=1}^{M}, $ a differentiable, closed-form solution for the affine transformation that aligns the landmarks can be obtained by minimizing the squared error:

\begin{equation}\
\label{AffineLoss}
    \varphi^*(\mathbf{X}_{mov}^{all}, \mathbf{X}_{fixed}^{all}) = \arg \min_{\varphi} \sum_{i=1}^{M} \Bigl\|\mathbf{B}^T \tilde{\mathbf{X}}_{mov}^{(i)} - \mathbf{X}_{fixed}^{(i)}\Bigr\|_F^2,
\end{equation}

\noindent where $\|\cdot\|_F$  define the Frobenius norm. By differentiating Eq.~\ref{AffineLoss} with respect to $\mathbf{B}$ and equating the result to zero, we obtain a closed-form solution, as derived in \cite{evan2022keymorph,wang2023robust}:

\begin{equation}
    \varphi^* = \mathbf{X}_{fixed}^{all} \, (\tilde{\mathbf{X}}_{mov}^{all})^T \Bigl( \tilde{\mathbf{X}}_{mov}^{all} \, (\tilde{\mathbf{X}}_{mov}^{all})^T \Big)^{-1}.
\end{equation}

\noindent
Note that for both rigid and affine transformations, the solutions correspond to a highly constrained system, meaning the landmarks are not perfectly matched in practice due to the constraints of these transformations. This limitation can be mitigated by using a transformation family with additional degrees of freedom.

\subsubsection{Thin-Plate Spline (TPS)}

The TPS transformation provides a non-rigid, parametric model for mapping coordinates. It yields a closed-form solution for interpolating corresponding landmarks \cite{donato2002approximate,bookstein1989principal,rohr2001landmark,zhao2022thin} and inherently includes affine mappings as a special case.

In this formulation, the TPS model is defined by
\begin{equation}
    T_\varphi(\mathbf{X}_{mov}) = \mathbf{B}^T \tilde{\mathbf{X}}_{mov} + \sum_{i=1}^{M} \mathbf{w}_i \, \psi\Bigl(\bigl\|\mathbf{X}_{mov}^{(i)} - \mathbf{X}_{mov}\bigr\|^2\Bigr),
\end{equation}
where \(\mathbf{B} \in \mathbb{R}^{D \times (D+1)}\) and each \(\mathbf{w}_i \in \mathbb{R}^D\) form the parameter set \(\varphi = \{\mathbf{B}, \mathbf{W}\}\) (with \(\mathbf{W} = \{\mathbf{w}_i\}_{i=1}^{M}\)). Also, $\psi(s) = s^{2} \ln(s)$. 

The transformation is designed to minimize the bending energy,

\begin{equation}
    E_T = \int_{\mathbb{R}^D} \Bigl\|\nabla^2 T(\mathbf{X}_{mov})\Bigr\|_F^2 \, d\mathbf{X}_{mov},
\end{equation}
which ensures that \(T\) is smooth with square-integrable second-order derivatives. The interpolation condition $T_\varphi\Bigl(\mathbf{X}_{mov}^{(i)}\Bigr) = \mathbf{X}_{fixed}^{(i)} $ is imposed along with the constraints

\begin{equation}
    \sum_{i=1}^{M} \mathbf{w}_i = \mathbf{0} \quad \text{and} \quad \sum_{i=1}^{M} \mathbf{w}_i \, \Bigl(\mathbf{X}_{mov}^{(i)}\Bigr)^T = \mathbf{0}.
\end{equation}

\noindent
These conditions lead to the following linear system for determining the parameters:
\begin{equation}
    \label{LinSysNew}
    \begin{bmatrix} 
        \mathbf{K} & \mathbf{H} \\[1mm]
        \mathbf{H}^T & \mathbf{0}
    \end{bmatrix}
    \begin{bmatrix} 
        \mathbf{W} \\[1mm]
        \mathbf{B}
    \end{bmatrix}
    =
    \begin{bmatrix} 
        \mathbf{X}_{fixed}^{all} \\[1mm]
        \mathbf{0}
    \end{bmatrix},
\end{equation}
where the matrix \(\mathbf{K} \in \mathbb{R}^{M \times M}\) has entries 
\[
K_{ij} = \psi\Bigl(\bigl\|\mathbf{X}_{mov}^{(i)} - \mathbf{X}_{mov}^{(j)}\bigr\|^2\Bigr),
\]
and \(\mathbf{H} \in \mathbb{R}^{M \times (D+1)}\) is constructed such that the \(i\)-th row is \(\tilde{\mathbf{X}}_{mov}^{(i)T}\).

\noindent
Thus, the solution for the optimal parameters is given by
\begin{equation}
    \varphi^* =
    \begin{bmatrix} 
        \mathbf{W}^* \\[1mm]
        \mathbf{B}^*
    \end{bmatrix}
    =
    \begin{bmatrix} 
        \mathbf{K} & \mathbf{H} \\[1mm]
        \mathbf{H}^T & \mathbf{0}
    \end{bmatrix}^{-1}
    \begin{bmatrix} 
        \mathbf{X}_{fixed}^{all} \\[1mm]
        \mathbf{0}
    \end{bmatrix}.
\end{equation}

\noindent
Finally, to handle potential noise in the data, the general TPS model can be extended by including a regularization term:
\begin{equation}
    \varphi^* = \arg \min_{\varphi} \sum_{i=1}^{M} \Bigl\|T_\varphi\Bigl(\mathbf{X}_{mov}^{(i)}\Bigr) - \mathbf{X}_{fixed}^{(i)}\Bigr\|^2 + \lambda \, I,
\end{equation}
where \(\lambda \) is a positive hyperparameter that controls the level of regularization. When \(\lambda \rightarrow \infty\), the optimal transformation tends toward an affine mapping. This behavior can be enforced by modifying the matrix \(\mathbf{K}\) to \(\mathbf{K} + \lambda \mathbf{I}\) in Eq. \eqref{LinSysNew}. Hence, the hyperparameter \(\lambda\) balances the trade-off between a fully nonlinear deformation (when \(\lambda\) is small) and an affine transformation (when \(\lambda\) is large).

\section{Results}

\begin{table*}[h]
  \centering
   \adjustbox{addcode={\begin{minipage}{\width}}{\caption{
   Subject-level quantitative results for Dice Score (\%) and Hausdorff Distance (mm) (mean$\pm$std) across all registration methods on the volunteer dataset. $*$ indicates statistically significant differences (p-value < 0.05) compared to the second-best method.
      }\end{minipage}},rotate=90,center}{
  \renewcommand{\arraystretch}{2.5}
  \resizebox{\textheight}{!}{
   \Huge
    \begin{tabular}{|c|c|c|c|c|c|c|c|c|c|c|c|c|c|c|c|c|c|}
    \hline
    \rowcolor{black}
    \Huge \textcolor{white}{Subject} &
    \Huge \textcolor{white}{Site} &
    \multicolumn{8}{c|}{\Huge \textcolor{white}{Dice Score (\%) \textcolor{green}{$\uparrow$}}} &
    \multicolumn{8}{c|}{\Huge \textcolor{white}{Hausdorff Distance (mm) \textcolor{green}{$\downarrow$}}} \\
    \hline
    \rowcolor{black}
    \Huge \textcolor{white}{ID} &
    \Huge \textcolor{white}{Type} &
    \Huge \textcolor{white}{NiftyReg} & \Huge \textcolor{white}{VoxelMorph} &
    \Huge \textcolor{white}{ANTS (Rigid)} & \Huge \textcolor{white}{ANTS (Affine)} &
    \Huge \textcolor{white}{ANTS (SyN)} &
    \Huge \textcolor{white}{DINOMotion (Rigid)} & \Huge \textcolor{white}{DINOMotion (Affine)} &
    \Huge \textcolor{white}{DINOMotion (Nonlinear)} &
    \Huge \textcolor{white}{NiftyReg} & \Huge \textcolor{white}{VoxelMorph} &
    \Huge \textcolor{white}{ANTS (Rigid)} & \Huge \textcolor{white}{ANTS (Affine)} &
    \Huge \textcolor{white}{ANTS (SyN)} &
    \Huge \textcolor{white}{DINOMotion (Rigid)} & \Huge \textcolor{white}{DINOMotion (Affine)} &
    \Huge \textcolor{white}{DINOMotion (Nonlinear)} \\
    \hline
    \Huge V1 & \Huge Pancreas &
    \Huge 48.34 $\pm$ 37.71 & 
    \Huge 41.93 $\pm$ 37.41 &
    \Huge 39.88 $\pm$ 34.93 & 
    \Huge 47.63 $\pm$ 33.77 &
    \Huge 53.54 $\pm$ 35.92 &
    \Huge 51.73 $\pm$ 27.53 & 
    \Huge 53.63 $\pm$ 27.27 &
    \Huge \textbf{61.93}\bigasterisk $\pm$ 24.56 &
    \Huge 14.95 $\pm$ 11.44 & 
    \Huge 16.18 $\pm$ 12.06 &
    \Huge 17.03 $\pm$ 11.80 & 
    \Huge 14.73 $\pm$ 10.54 &
    \Huge 13.66 $\pm$ 11.03 &
    \Huge 12.74 $\pm$  9.39 & 
    \Huge 12.43 $\pm$  9.37 &
    \Huge \textbf{10.86}\bigasterisk $\pm$  8.58\\
    
    \Huge V2 & \Huge Liver &
    \Huge \textbf{95.20} $\pm$  0.98 & 
    \Huge 91.99 $\pm$  1.41 &
    \Huge 89.88 $\pm$  1.30 & 
    \Huge 90.27 $\pm$  0.78 &
    \Huge 94.57 $\pm$  0.65 &
    \Huge 92.50 $\pm$  0.69 & 
    \Huge 92.25 $\pm$  0.49 &
    \Huge 92.78 $\pm$  0.59 &
    \Huge  7.32 $\pm$  2.08 & 
    \Huge  8.54 $\pm$  1.17 &
    \Huge  8.26 $\pm$  0.92 & 
    \Huge  8.05 $\pm$  0.72 &
    \Huge  6.15 $\pm$  1.30 &
    \Huge  \textbf{6.14} $\pm$  1.12 & 
    \Huge  6.41 $\pm$  0.99 &
    \Huge  7.10 $\pm$  1.05 \\
    
    \Huge V3 & \Huge Kidney &
    \Huge 79.45 $\pm$ 19.55 & 
    \Huge 79.04 $\pm$ 18.52 &
    \Huge 78.16 $\pm$ 15.64 & 
    \Huge 77.74 $\pm$ 16.66 &
    \Huge 81.56 $\pm$ 16.58 &
    \Huge 88.58 $\pm$  7.85 & 
    \Huge 88.30 $\pm$  8.13 &
    \Huge \textbf{91.17}\bigasterisk $\pm$  6.10 &
    \Huge 11.46 $\pm$  8.95 & 
    \Huge 12.01 $\pm$  8.67 &
    \Huge 11.43 $\pm$  7.31 & 
    \Huge 11.48 $\pm$  7.58 &
    \Huge 10.32 $\pm$  7.77 &
    \Huge  6.85 $\pm$  4.61 & 
    \Huge  6.97 $\pm$  4.76 &
    \Huge  \textbf{6.20}\bigasterisk $\pm$  4.08 \\
    
    \Huge V4 & \Huge Lung &
    \Huge 97.19 $\pm$  1.19 & 
    \Huge 97.46 $\pm$  1.58 &
    \Huge 96.06 $\pm$  1.76 & 
    \Huge 96.52 $\pm$  1.51 &
    \Huge 97.17 $\pm$  0.44 &
    \Huge 96.50 $\pm$  1.77 & 
    \Huge 96.71 $\pm$  1.20 &
    \Huge \textbf{97.97} $\pm$  0.85 &
    \Huge  4.75 $\pm$  3.18 & 
    \Huge  5.64 $\pm$  3.80 &
    \Huge  5.53 $\pm$  3.20 & 
    \Huge  5.39 $\pm$  3.35 &
    \Huge  4.40 $\pm$  2.61 &
    \Huge  4.73 $\pm$  2.92 & 
    \Huge  4.91 $\pm$  2.99 &
    \Huge  \textbf{4.09}\bigasterisk $\pm$  3.14 \\
    
    \Huge V5 & \Huge Lung &
    \Huge \textbf{93.46 $\pm$  2.98} & 
    \Huge 92.75 $\pm$  3.15 &
    \Huge 91.95 $\pm$  3.02 & 
    \Huge 92.04 $\pm$  2.87 &
    \Huge 93.86 $\pm$  2.50 &
    \Huge 91.76 $\pm$  3.48 & 
    \Huge 91.85 $\pm$  3.30 &
    \Huge 92.32 $\pm$  3.09 &
    \Huge 10.47 $\pm$  6.10 & 
    \Huge 10.67 $\pm$  5.82 &
    \Huge 10.41 $\pm$  5.33 & 
    \Huge  9.96 $\pm$  5.07 &
    \Huge  9.00 $\pm$  4.96 &
    \Huge  9.27 $\pm$  4.08 & 
    \Huge  9.31 $\pm$  4.23 &
    \Huge  \textbf{9.00}\bigasterisk $\pm$  4.03 \\
    
    \Huge V6 & \Huge Pancreas &
    \Huge 52.58 $\pm$ 35.18 & 
    \Huge 42.58 $\pm$ 31.63 &
    \Huge 40.68 $\pm$ 28.28 & 
    \Huge 47.21 $\pm$ 30.63 &
    \Huge 55.17 $\pm$ 32.52 &
    \Huge 53.49 $\pm$ 32.77 & 
    \Huge 53.65 $\pm$ 32.73 &
    \Huge \textbf{55.24}\bigasterisk $\pm$ 33.25 &
    \Huge 11.46 $\pm$ 10.86 & 
    \Huge 12.45 $\pm$  9.75 &
    \Huge 13.26 $\pm$  9.60 & 
    \Huge 12.48 $\pm$  9.75 &
    \Huge 11.12 $\pm$ 10.12 &
    \Huge 10.27 $\pm$ 10.23 & 
    \Huge 10.26 $\pm$ 10.24 &
    \Huge \textbf{10.06}\bigasterisk $\pm$ 10.27 \\
    
    \Huge V7 & \Huge Lung &
    \Huge \textbf{97.63} $\pm$  0.34 & 
    \Huge 91.41 $\pm$  2.80 &
    \Huge 90.63 $\pm$  3.43 & 
    \Huge 91.93 $\pm$  3.09 &
    \Huge 97.29 $\pm$  0.48 &
    \Huge 92.05 $\pm$  2.82 & 
    \Huge 95.53 $\pm$  1.01 &
    \Huge 96.65 $\pm$  0.42 &
    \Huge  5.82 $\pm$  1.15 & 
    \Huge  9.72 $\pm$  1.91 &
    \Huge  9.19 $\pm$  2.19 & 
    \Huge  8.95 $\pm$  2.20 &
    \Huge \textbf{5.15} $\pm$  1.58 &
    \Huge  8.14 $\pm$  1.69 & 
    \Huge  6.65 $\pm$  1.89 &
    \Huge  6.53 $\pm$  1.72 \\
    
    \Huge V8 & \Huge Kidney &
    \Huge 93.33 $\pm$  4.52 &
    \Huge 95.30 $\pm$  1.83 &
    \Huge 92.93 $\pm$  3.46 & 
    \Huge 92.95 $\pm$  3.56 &
    \Huge 94.40 $\pm$  2.49 &
    \Huge 93.60 $\pm$  2.71 & 
    \Huge 93.28 $\pm$  3.19 &
    \Huge \textbf{96.07}\bigasterisk $\pm$  1.57 & 
    \Huge  4.45 $\pm$  2.20&
    \Huge  4.03 $\pm$  2.11 &
    \Huge  4.73 $\pm$  1.80 & 
    \Huge  4.73 $\pm$  1.83 &
    \Huge  4.17 $\pm$  1.75 &
    \Huge  4.53 $\pm$  2.25 & 
    \Huge  4.59 $\pm$  2.18 &
    \Huge  \textbf{3.52}\bigasterisk $\pm$  2.23\\

    \Huge V9 & \Huge Prostate &
    \Huge 87.10 $\pm$  9.26 & 
    \Huge 87.98 $\pm$  5.89 &
    \Huge 82.51 $\pm$  7.50 & 
    \Huge 82.88 $\pm$  7.55 &
    \Huge  \textbf{90.20} $\pm$  3.44 &
    \Huge 83.37 $\pm$  8.81 & 
    \Huge 83.19 $\pm$  7.27 &
    \Huge 82.76 $\pm$  7.38 &
    \Huge  3.88 $\pm$  2.31 & 
    \Huge  3.54 $\pm$  1.52 &
    \Huge  4.34 $\pm$  1.20 & 
    \Huge  4.30 $\pm$  1.19 &
    \Huge  \textbf{3.04} $\pm$  0.93 &
    \Huge  3.59 $\pm$  1.32 & 
    \Huge  3.58 $\pm$  1.16 &
    \Huge  3.76 $\pm$  1.09 \\
    
    \Huge V10 & \Huge Liver &
    \Huge 85.47 $\pm$ 15.90 & 
    \Huge 81.52 $\pm$ 18.50 &
    \Huge 79.77 $\pm$ 18.27 & 
    \Huge 80.53 $\pm$ 17.58 &
    \Huge 84.50 $\pm$ 17.35 &
    \Huge 87.33 $\pm$ 12.80 & 
    \Huge 87.69 $\pm$ 12.50 &
    \Huge \textbf{89.99}\bigasterisk $\pm$  8.58 &
    \Huge 10.49 $\pm$  9.09 & 
    \Huge 11.40 $\pm$ 10.13 &
    \Huge 11.47 $\pm$  9.85 & 
    \Huge 11.19 $\pm$  9.54 &
    \Huge 10.07 $\pm$  9.65 &
    \Huge  \textbf{8.62}\bigasterisk $\pm$  7.61 & 
    \Huge  9.49 $\pm$  8.94 &
    \Huge  8.91 $\pm$  8.38 \\
    \hline
    \end{tabular}
  }
  }
  \label{table1}
\end{table*}

\begin{table*}[tb]
  \centering
  \adjustbox{addcode={\begin{minipage}{\width}}{\caption{Patient-level quantitative results for Dice Score (\%) and Hausdorff Distance (mm) (mean$\pm$std) across all registration methods on the patient dataset. $*$ indicates statistically significant differences (p-value < 0.05) compared to the second-best method.
      }\end{minipage}},rotate=90,center}{
  \renewcommand{\arraystretch}{2.5}
  \resizebox{\textheight }{!}{
   \Huge
    \begin{tabular}{|c|c|c|c|c|c|c|c|c|c|c|c|c|c|c|c|c|c|}
    \hline
    \rowcolor{black}
    \Huge \textcolor{white}{Patient} &
    \Huge \textcolor{white}{Site} &
    \multicolumn{8}{c|}{\Huge \textcolor{white}{Dice Score (\%) \textcolor{green}{$\uparrow$}}} &
    \multicolumn{8}{c|}{\Huge \textcolor{white}{Hausdorff Distance (mm) \textcolor{green}{$\downarrow$}}} \\
    \hline
    \rowcolor{black}
    \Huge \textcolor{white}{ID} &
    \Huge \textcolor{white}{Type} &
    \Huge \textcolor{white}{NiftyReg} & 
    \Huge \textcolor{white}{VoxelMorph} &
    \Huge \textcolor{white}{ANTS (Rigid)} & 
    \Huge \textcolor{white}{ANTS (Affine)} &
    \Huge \textcolor{white}{ANTS (SyN)} &
    \Huge \textcolor{white}{DINOMotion (Rigid)} & 
    \Huge \textcolor{white}{DINOMotion (Affine)} &
    \Huge \textcolor{white}{DINOMotion (Nonlinear)} &
    
    \Huge \textcolor{white}{NiftyReg} & 
    \Huge \textcolor{white}{VoxelMorph} &
    \Huge \textcolor{white}{ANTS (Rigid)} &
    \Huge \textcolor{white}{ANTS (Affine)} &
    \Huge \textcolor{white}{ANTS (SyN)} &
    \Huge \textcolor{white}{DINOMotion (Rigid)} &
    \Huge \textcolor{white}{DINOMotion (Affine)} &
    \Huge \textcolor{white}{DINOMotion (Nonlinear)} \\
    \hline
    
    \Huge 1  & \Huge Kidney   & 
    \Huge 85.96 ± 6.74& 
    \Huge 88.37 ± 4.62& 
    \Huge 86.78 ± 5.95& 
    \Huge 89.45 ± 5.65& 
    \Huge \textbf{92.13} ± 4.58& 
    \Huge 90.65 ± 8.16&
    \Huge 90.77 ± 8.25&
    \Huge 91.57 ± 3.70&

    \Huge 6.35 ± 2.06&
    \Huge 5.95 ± 2.58& 
    \Huge 5.74 ± 1.84& 
    \Huge 5.56 ± 2.01& 
    \Huge 4.96 ± 1.59&
    \Huge 5.34 ± 2.12&
    \Huge 5.24 ± 2.15&
    \Huge \textbf{4.93} ± 2.00\\

    \Huge 2  & 
    \Huge Liver  
    & \Huge 84.16 ± 17.01
    & \Huge 93.67 ± 3.79
    & \Huge 85.90 ± 12.18
    & \Huge 91.76 ± 6.83
    & \Huge 95.04 ± 3.25
    & \Huge 91.45 ± 6.84
    & \Huge 91.99 ± 3.91
    & \Huge \textbf{95.23} ± 3.22

    & \Huge 12.57 ± 10.54
    & \Huge 7.82 ± 6.08
    & \Huge 11.78 ± 9.00
    & \Huge 8.05 ± 6.61
    & \Huge 7.02 ± 7.11
    & \Huge 8.04 ± 7.03
    & \Huge 8.03 ± 7.02
    & \Huge \textbf{6.92} ± 6.15\\

    \Huge 3 & \Huge Kidney   
    & \Huge 70.63 ± 10.85
    & \Huge 75.91 ± 11.69
    & \Huge 72.52 ± 15.01
    & \Huge 74.69 ± 14.58
    & \Huge \textbf{81.26} ± 10.55
    & \Huge 79.04 ± 9.37
    & \Huge 80.78 ± 14.22
    & \Huge 81.09 ± 14.29

    & \Huge 5.37 ± 2.25
    & \Huge 6.02 ± 3.57
    & \Huge 5.42 ± 2.80
    & \Huge 4.69 ± 2.22
    & \Huge 4.34 ± 2.58
    & \Huge 4.18 ± 2.34
    & \Huge 4.14 ± 2.25
    & \Huge \textbf{4.04}\bigasterisk ± 2.36  \\

    \Huge 4  & \Huge Liver      
    &\Huge 89.94 ± 8.47
    &\Huge 91.93 ± 7.98
    &\Huge 89.24 ± 10.13
    &\Huge 89.14 ± 10.20
    &\Huge 93.34 ± 7.16
    &\Huge 90.65 ± 8.64
    &\Huge 90.69 ± 8.77
    &\Huge \textbf{94.79}\bigasterisk ± 5.10

    &\Huge 5.08 ± 2.45
    &\Huge 5.56 ± 2.90
    &\Huge 5.47 ± 3.12
    &\Huge 5.48 ± 3.09
    &\Huge 4.76 ± 2.52
    &\Huge 4.76 ± 2.59 
    &\Huge 4.75 ± 2.62 
    &\Huge \textbf{3.93}\bigasterisk ± 2.13 \\

    \Huge 5  & \Huge Liver     
    &\Huge 78.21 ± 14.21
    &\Huge 77.57 ± 13.79
    &\Huge 71.15 ± 15.74
    &\Huge 72.29 ± 14.98
    &\Huge 77.69 ± 16.00
    &\Huge 82.96 ± 8.58
    &\Huge 84.19 ± 8.23
    &\Huge \textbf{89.73}\bigasterisk ± 4.62

    &\Huge 14.90 ± 9.83
    &\Huge 14.40 ± 8.50
    &\Huge 14.77 ± 8.44
    &\Huge 14.60 ± 8.39
    &\Huge 13.29 ± 8.97
    &\Huge 10.76 ± 6.33
    &\Huge 10.76 ± 6.55
    &\Huge \textbf{8.10}\bigasterisk ± 4.78\\

    \Huge 6  & \Huge Liver    
    & \Huge 79.97 ± 7.39
    & \Huge 85.02 ± 10.73
    & \Huge 82.16 ± 12.24
    & \Huge 82.92 ± 11.21
    & \Huge 87.27 ± 11.04
    & \Huge 82.68 ± 5.23
    & \Huge 82.98 ± 6.22
    & \Huge \textbf{87.70}\bigasterisk ± 10.63

    & \Huge 11.23 ± 2.97
    & \Huge 11.51 ± 5.46
    & \Huge 11.36 ± 5.49
    & \Huge 11.07 ± 5.24
    & \Huge 10.67 ± 5.16
    & \Huge 10.76 ± 2.96
    & \Huge 10.27 ± 3.49
    & \Huge \textbf{9.72}\bigasterisk ± 5.17\\

    \Huge 7  & \Huge Kidney  
    & \Huge 92.83 ± 5.36
    & \Huge 94.19 ± 4.07
    & \Huge 90.74 ± 8.79
    & \Huge 90.79 ± 8.80
    & \Huge \textbf{94.74} ± 4.81
    & \Huge 91.89 ± 6.73
    & \Huge 92.11 ± 6.16
    & \Huge 94.67 ± 4.52

    & \Huge 4.69 ± 1.45
    & \Huge 4.09 ± 1.71
    & \Huge 5.62 ± 2.53
    & \Huge 5.61± 2.53
    & \Huge 3.94 ± 1.22
    & \Huge 5.38 ± 1.87
    & \Huge 4.99 ± 1.84
    & \Huge \textbf{3.82} ± 1.13\\

    \Huge 8  & \Huge Liver    
    & \Huge 96.66 ± 1.59
    & \Huge 95.64 ± 1.83
    & \Huge 93.95 ± 4.23
    & \Huge 94.09 ± 4.14
    & \Huge 96.59 ± 1.65
    & \Huge 95.30 ± 2.47 
    & \Huge 95.02 ± 2.71
    & \Huge \textbf{96.67} ± 1.32

    & \Huge 4.00 ± 2.44
    & \Huge 4.14 ± 1.83
    & \Huge 4.41 ± 2.19 
    & \Huge 4.32 ± 2.18 
    & \Huge 3.12 ± 1.14
    & \Huge 3.70 ± 1.41
    & \Huge 3.66 ± 1.37
    & \Huge \textbf{3.08} ± 0.76 \\

    \Huge 9  & \Huge Kidney
    &\Huge 70.64 ± 18.89
    &\Huge 70.65 ± 19.14
    &\Huge 69.67 ± 17.99
    &\Huge 69.80 ± 17.65
    &\Huge \textbf{75.37} ± 19.23
    &\Huge 70.40 ± 18.22
    &\Huge 72.10 ± 20.30
    &\Huge 74.58 ± 15.08
    
    &\Huge 14.88 ± 11.71
    &\Huge 14.98 ± 10.51
    &\Huge 14.88 ± 10.33
    &\Huge 14.66 ± 10.25
    &\Huge 14.30 ± 10.48
    &\Huge 13.53 ± 9.98
    &\Huge \textbf{12.38}\bigasterisk ± 8.63
    &\Huge 13.87 ± 10.38\\

    \Huge 10 & \Huge Liver    
    & \Huge 86.90 ± 7.03
    & \Huge 92.80 ± 3.12
    & \Huge 87.23 ± 8.76
    & \Huge 91.48 ± 5.63
    & \Huge 96.04 ± 1.22
    & \Huge 92.98 ± 4.48 
    & \Huge 93.75 ± 3.59 
    & \Huge \textbf{96.09} ± 0.82
    
    & \Huge 8.18 ± 3.97
    & \Huge 4.96 ± 2.04
    & \Huge 6.77 ± 2.78 
    & \Huge 5.15 ± 1.46
    & \Huge 3.51 ± 1.53
    & \Huge 4.66 ± 1.76
    & \Huge 4.44 ± 1.63
    & \Huge \textbf{3.34} ± 1.59\\

    \Huge 11 & \Huge Pancreas 
    & \Huge 83.83 ± 3.46
    & \Huge 81.02 ± 5.06
    & \Huge 83.11 ± 7.36
    & \Huge 84.28 ± 6.01
    & \Huge 86.61 ± 6.71
    & \Huge 83.57 ± 5.29
    & \Huge 85.05 ± 4.51
    & \Huge \textbf{88.34}\bigasterisk ± 2.39
    
    & \Huge 4.46 ± 1.39
    & \Huge 6.29 ± 2.07
    & \Huge 5.35 ± 1.16
    & \Huge 5.28 ± 1.07
    & \Huge 5.42 ± 1.74
    & \Huge 5.24 ± 1.08
    & \Huge 5.18 ± 1.08
    & \Huge \textbf{4.34}\bigasterisk  ± 0.60\\

    \Huge 12 & \Huge Pancreas 
    & \Huge 84.31 ± 4.07
    & \Huge 80.70 ± 10.06
    & \Huge 80.93 ± 10.12
    & \Huge 83.11 ± 8.40
    & \Huge 85.25 ± 8.51
    & \Huge 82.17 ± 6.38
    & \Huge 84.90 ± 5.31
    & \Huge \textbf{87.52}\bigasterisk ± 3.75
    
    & \Huge 5.03 ± 3.15
    & \Huge 5.55 ± 2.72
    & \Huge 5.28 ± 2.09 
    & \Huge 4.94 ± 2.29
    & \Huge 4.11 ± 1.70
    & \Huge 4.96 ± 1.96
    & \Huge 4.58 ± 1.69 
    & \Huge \textbf{3.99}\bigasterisk ± 1.43\\

    \Huge 13 & \Huge Pancreas 
    & \Huge 82.50 ± 7.62
    & \Huge 80.60 ± 6.93 
    & \Huge 79.35 ± 8.48
    & \Huge 84.79 ± 4.82
    & \Huge 87.42 ± 5.09
    & \Huge 84.80 ± 6.04
    & \Huge 86.25 ± 5.21
    & \Huge \textbf{87.69} ± 4.55
    
    & \Huge 8.75 ± 3.15
    & \Huge 9.46 ± 3.23
    & \Huge 8.15 ± 3.52
    & \Huge 7.76 ± 3.55
    & \Huge 7.57 ± 2.38
    & \Huge 8.52 ± 3.18
    & \Huge 7.58 ± 2.99
    & \Huge \textbf{7.35} ± 2.59\\

    \Huge 14 & \Huge Pancreas 
    & \Huge 87.78 ± 4.51
    & \Huge 86.94 ± 3.83
    & \Huge \textbf{88.41} ± 5.35
    & \Huge \textbf{88.41} ± 5.35
    & \Huge 86.25 ± 5.36 
    & \Huge 86.96 ± 6.52
    & \Huge 85.83 ± 6.74
    & \Huge 87.89 ± 4.79
    
    & \Huge 7.54 ± 2.36
    & \Huge 7.34 ± 3.19
    & \Huge 6.52 ± 4.09
    & \Huge 6.52 ± 4.09
    & \Huge 7.35 ± 3.39
    & \Huge \textbf{6.26}\bigasterisk ± 3.35
    & \Huge 7.09 ± 3.56
    & \Huge 7.16 ± 3.99\\

    \Huge 15 & \Huge Liver 
    & \Huge 84.90 ± 16.78
    & \Huge 90.96 ± 4.10
    & \Huge 86.18 ± 14.62
    & \Huge 88.66 ± 6.76
    & \Huge 94.10 ± 1.58
    & \Huge 90.12 ± 5.16
    & \Huge 90.89 ± 4.21
    & \Huge \textbf{94.31} ± 1.74
    
    & \Huge 8.18 ± 3.43
    & \Huge 6.63 ± 2.42
    & \Huge 8.35 ± 3.89
    & \Huge 7.25 ± 2.98
    & \Huge 5.76 ± 2.70
    & \Huge 6.83 ± 2.75
    & \Huge 6.51 ± 2.60
    & \Huge \textbf{5.54} ± 2.81\\

    \Huge 16 & \Huge Pancreas 
    & \Huge 73.65 ± 23.14
    & \Huge 71.00 ± 20.91
    & \Huge 55.47 ± 27.36
    & \Huge 56.09 ± 21.90
    & \Huge 77.63 ± 22.09
    & \Huge 68.95 ± 21.62
    & \Huge 71.10 ± 22.62
    & \Huge \textbf{80.00} ± 18.65
    
    & \Huge 6.35 ± 7.84
    & \Huge 9.25 ± 9.71
    & \Huge 12.88 ± 9.63
    & \Huge 10.65 ± 7.62
    & \Huge 8.28 ± 9.04
    & \Huge 8.85 ± 7.78
    & \Huge 8.45 ± 8.01
    & \Huge \textbf{6.66} \bigasterisk ± 6.98\\

    \Huge 17 & \Huge Pancreas 
    & \Huge 67.97 ± 17.60
    & \Huge 60.64 ± 10.85
    & \Huge 56.45 ± 14.67
    & \Huge 60.40 ± 13.60
    & \Huge 72.66 ± 19.67
    & \Huge 66.41 ± 18.04
    & \Huge 67.79 ± 18.29
    & \Huge \textbf{72.95}  ± 17.91
    
    & \Huge 12.79 ± 6.90
    & \Huge 12.30 ± 6.63
    & \Huge 14.15 ± 6.24
    & \Huge 12.94 ± 6.99
    & \Huge 11.76 ± 5.97
    & \Huge 12.43 ± 5.95
    & \Huge 12.17 ± 6.22
    & \Huge \textbf{10.94} \bigasterisk  ± 5.74
    \\

    \Huge 18 & \Huge Pancreas 
    & \Huge 82.23 ± 6.90
    & \Huge 77.79 ± 9.47
    & \Huge 75.31 ± 13.02
    & \Huge 77.51 ± 11.45
    & \Huge 87.01 ± 4.40
    & \Huge 81.80 ± 6.37
    & \Huge 81.29 ± 7.78
    & \Huge \textbf{89.53}\bigasterisk ± 3.35
    
    & \Huge 5.83 ± 1.68
    & \Huge 6.82 ± 2.54
    & \Huge 6.75 ± 1.51
    & \Huge 6.14 ± 1.72
    & \Huge 4.44 ± 0.95
    & \Huge 5.48 ± 1.09
    & \Huge 5.32 ± 0.97
    & \Huge \textbf{3.88}\bigasterisk  ± 0.62\\

    \Huge 19 & \Huge Pancreas 
    & \Huge 87.18 ± 5.51
    & \Huge 84.36 ± 6.64
    & \Huge 80.76 ± 10.41
    & \Huge 84.78 ± 6.32
    & \Huge 88.59 ± 5.07
    & \Huge 83.97 ± 3.54
    & \Huge 84.51 ± 3.39
    & \Huge \textbf{91.35}\bigasterisk ± 1.32
    
    & \Huge 4.12 ± 2.00
    & \Huge 4.85 ± 2.98
    & \Huge 4.77 ± 1.96
    & \Huge 4.45 ± 1.36
    & \Huge 4.42 ± 2.70
    & \Huge 4.64 ± 0.92
    & \Huge 4.64 ± 0.92
    & \Huge \textbf{3.11}\bigasterisk  ± 0.96\\

    \Huge 20 & \Huge Liver    
    &\Huge 92.12 ± 3.99
    &\Huge 88.65 ± 7.03
    &\Huge 91.20 ± 6.10
    &\Huge 92.60 ± 4.52
    &\Huge 95.81 ± 1.48
    &\Huge 92.94 ± 3.90
    &\Huge 93.60 ± 2.78 
    &\Huge \textbf{96.32}\bigasterisk ± 1.20

    &\Huge  4.75 ± 1.80
    &\Huge  5.11 ± 2.18
    &\Huge 4.17 ± 1.90
    &\Huge  3.91 ± 1.97
    &\Huge  3.76 ± 2.66
    &\Huge  3.67 ± 1.83
    &\Huge 3.57 ± 1.99
    &\Huge \textbf{2.78} \bigasterisk ± 1.55\\

    \Huge 21 & \Huge Pancreas & 
    
    \Huge 85.89 ± 4.69& 
    \Huge  88.75 ± 5.47& 
    \Huge 86.46 ± 7.82& 
    \Huge 87.18 ± 7.70& 
    \Huge 92.83 ± 3.78& 
    \Huge 87.57 ± 5.62& 
    \Huge 88.49 ± 5.34& 
    \Huge \textbf{93.91}  ± 1.61&

    \Huge 5.62 ± 1.73&
    \Huge 6.84 ± 2.92& 
    \Huge 5.77 ± 2.51& 
    \Huge 5.67 ± 2.53& 
    \Huge 4.16 ± 1.35& 
    \Huge 5.56 ± 1.97& 
    \Huge 5.05 ± 1.57& 
    \Huge \textbf{4.05}  ± 1.72\\

    \Huge 22 &
    \Huge Prostate &
    \Huge 45.43 ± 24.96& 
    \Huge 56.15 ± 35.29& 
    \Huge 32.55 ± 25.95&
    \Huge 33.87 ± 24.64&
    \Huge 54.35 ± 28.38&
    \Huge 45.69 ± 26.89&
    \Huge 51.77 ± 29.19 & 
    \Huge \textbf{58.84}\bigasterisk  ± 33.09 &
    
    \Huge 10.87 ± 8.78& 
    \Huge 6.78 ± 4.32&
    \Huge 7.92 ± 3.44&
    \Huge 7.88 ± 3.42 &
    \Huge 6.72 ± 3.90&
    \Huge 7.52 ± 5.73&
    \Huge 7.88 ± 3.42  &
    \Huge \textbf{6.27}\bigasterisk  ± 4.96\\

    \Huge 23 &
    \Huge Prostate &
    \Huge 47.97 ± 31.39&
    \Huge 78.10 ± 6.03&
    \Huge 56.15 ± 32.32&
    \Huge 58.05 ± 31.54 &
    \Huge 78.18 ± 10.69&
    \Huge 45.18 ± 30.28 &
    \Huge 45.41 ± 31.55 & 
    \Huge \textbf{85.22}  \bigasterisk  ± 3.29 & 
    
    \Huge 5.03 ± 2.42 & 
    \Huge 3.70 ± 1.41 & 
    \Huge 4.35 ± 2.52 & 
    \Huge 4.46 ± 1.07& 
    \Huge 4.14 ± 2.40 & 
    \Huge 5.03 ± 2.42 & 
    \Huge 4.86 ± 2.26 &  
    \Huge \textbf{2.83} \bigasterisk  ± 1.42\\
    \hline
    \end{tabular}
  }
   }
  \label{tablePatientLevelPatient}
\end{table*}

\bibliographystyle{IEEEtran}
\bibliography{ref.bib}